%% file: draftv0.tex
\documentclass[reprint,superscriptaddress,amsmath,amssymb,aps,prb]{revtex4-2}

\usepackage{graphicx}
\usepackage{dcolumn}
\usepackage{bm}
\usepackage{subcaption}

\newcommand{\icol}[1]{
  \left(\begin{smallmatrix}#1\end{smallmatrix}\begin{smallmatrix}#1\end{smallmatrix}\begin{smallmatrix}#1\end{smallmatrix}\begin{smallmatrix}#1\end{smallmatrix}\begin{smallmatrix}#1\end{smallmatrix}\begin{smallmatrix}#1\end{smallmatrix}\begin{smallmatrix}#1\end{smallmatrix}\begin{smallmatrix}#1\end{smallmatrix}\right)%
}

\usepackage[bookmarks=false,linkcolor=blue,urlcolor=blue,colorlinks,citecolor=magenta]{hyperref}

\begin{document}


\title{Flat bands and multi-state memory devices from chiral domain wall \\ superlattices in magnetic Weyl semimetals}


\author{Vivian Rogers}
 \email{vivian.rogers@utexas.edu}
 \affiliation{%
 Chandra Department of Electrical and Computer Engineering, The University of Texas at Austin, Austin TX 78712
}%
\author{Swati Chaudhary}%
 \email{swati.chaudhary@austin.utexas.edu}
 \affiliation{Department of Physics, The University of Texas at Austin, Austin, Texas 78712, USA}
\affiliation{Department of Physics, Northeastern University, Boston, Massachusetts 02115, USA}
\affiliation{Department of Physics, Massachusetts Institute of Technology, Cambridge, Massachusetts 02139, USA}
\author{Richard Nguyen}
\affiliation{%
 Chandra Department of Electrical and Computer Engineering, The University of Texas at Austin, Austin TX 78712
}%
\author{Jean Anne Incorvia}%
 \email{incorvia@austin.utexas.edu}
\affiliation{%
 Chandra Department of Electrical and Computer Engineering, The University of Texas at Austin, Austin TX 78712
}%


\begin{abstract}
We propose a novel analog memory device utilizing the gigantic magnetic Weyl semimetal (MWSM) domain wall (DW) magnetoresistance. We predict that the nucleation of domain walls between contacts will strongly modulate the conductance and allow for multiple memory states, which has been long sought-after for use in magnetic random access memories or memristive neuromorphic computing platforms. We motivate this conductance modulation by analyzing the electronic structure of the helically-magnetized MWSM Hamiltonian, and report tunable flat bands in the direction of transport in a helically-magnetized region of the sample for Bloch and N\'eel-type domain walls via the onset of a local axial Landau level spectrum within the bulk of the superlattice. We show that Bloch devices also provide means for the generation of chirality-polarized currents, which provides a path towards nanoelectronic utilization of chirality as a new degree of freedom in spintronics. 
\end{abstract}

\maketitle



\section{\label{sec:intro}Introduction}

Topological materials such as topological insulators, Dirac semimetals, and Weyl semimetals have gained much interest for their litany of novel electron transport behaviors and potential utility in nanoelectronic devices~\cite{armitage_weyl_2018,burkov2018weyl, ilan_pseudo-electromagnetic_2020}. Weyl semimetals in particular must break either inversion or time-reversal symmetry (TRS). In a magnetic Weyl semimetal (MWSM), a Dirac cone is split into two Weyl cones with opposite chiralities when TRS is broken, giving rise to effects such as the chiral anomaly or negative longitudinal magnetoresistance (MR)~\cite{hu2019transport,nagaosa2020transport}. 



In intrinsically magnetic Weyl semimetals, the helicity of Weyl fermions can be controlled by changing the magnetization of the material, giving rise to large longitudinal MR via the helicity mismatch of carriers in differently-magnetized portions of a sample. Previous works~\cite{deSousa2021,kobayashi_helicity-protected_2018} have looked at this MR associated with transport across MWSM domain walls (DWs) and a magnetic tunnel junction (MTJ) constructed from MWSMs, both of which conclude that on/off ratios could be improved significantly ($10^4\% \text{ vs } 10^2 \%$) over the MRs utilized in traditional CoFeB/MgO MTJs. This giant Weyl MR arising from spatially-varying magnetic textures is expected to be resilient to disorder and large when compared to the DW resistances attributed to anisotropic MR or spin-mistracking in a nontopological DW~\cite{kent_DWR_2001,roxy_DWM_TR_2020}. Such a material platform has been long-sought after in neuromorphic computing or analog memory architectures~\cite{chen_multiply_2021,li_memristor_review_2018, wan_compute_memory_2022, wang_multistate_memristors_2022, zarcone_analog_2020}, but the fast write times and energy efficiency of existing spintronic memories have been offset by inefficient readout owing to low MRs in MTJs, often necessitating peripheral CMOS circuitry~\cite{jung_crossbar_2022, brigner_purelyspintronic_2022}. 

\begin{figure}[]
\includegraphics[width=\linewidth]{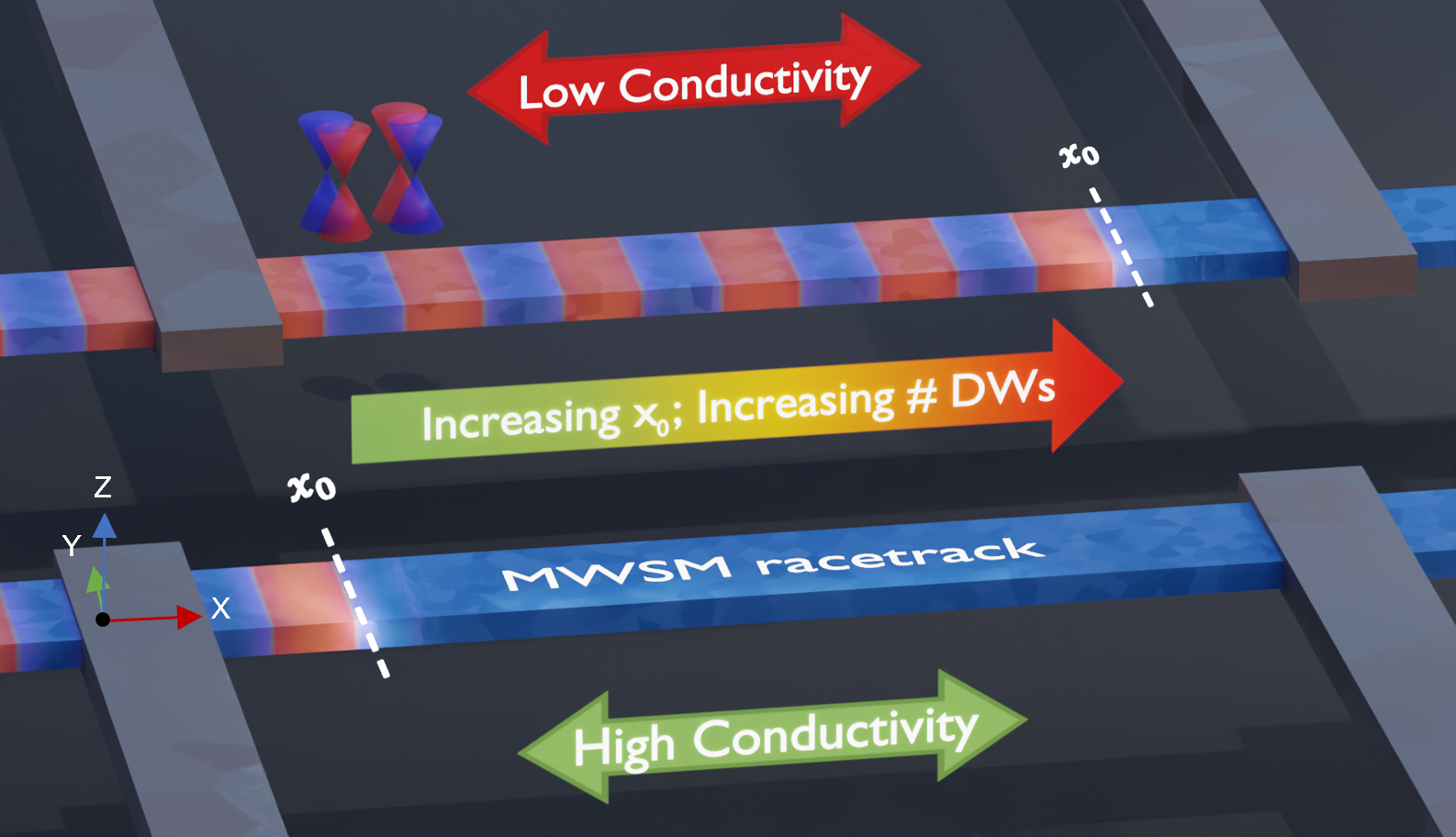}%
\caption{\label{fig:DWlat} Diagram of DW lattice MWSM device. $x_0$ represents the start position of a series of chiral DWs (red and blue domains depicted). In the bottom device, few MWSM DWs are present between nonmagnetic grey contacts. In the top device, $x_0$ is increased, injecting DWs into the active region, increasing resistance between the contacts. }
\end{figure}

Experimentally, an elevated longitudinal MR of $\approx7\%$ was observed by nucleating DWs in a sample of Weyl ferromagnet Co\textsubscript{3}Sn\textsubscript{2}S\textsubscript{2}~\cite{shiogai_co3sn2s2} at 100 K. Other works~\cite{araki_localized_2018, grushin_inhomogeneous_weyl_2016, kurebayashi_theory_2019} predict that the chirality-magnetization locking will cause carriers to localize at discontinuities in the magnetization profile, which along with the recently observed topological hall torque~\cite{yamanouchi_srruo3_THT_2022, wang_co3sn2s2_THT_2022}, opens up orders-of-magnitude more efficient means for electronic control of magnetic information in MWSMs. Recently, one helitronic magnetic memory platform has been proposed for unconventional computing, for which MWSM DW MRs could massively improve readout efficiencies~\cite{bechler_helitronics_2023}. Furthermore, the carrier localization and chirality - magnetization locking in MWSMs motivates interest in transport and the electronic structure of the periodic exchange superlattices, with the starting intuition that conductance should be forbidden in the direction of alternating magnetic texture. Previous works~\cite{araki_localized_2018, grushin_inhomogeneous_weyl_2016} have shown that N\'eel DWs will localize carriers in 1D coplanar to the DW, but a detailed understanding of the connection between periodic magnetic textures, artificial gauge fields, and spinful electronic bandstructures in 3D is lacking.

In this work, we propose a novel multi-state MWSM DW memory device (depicted in Fig~\ref{fig:DWlat}) and investigate the ensuing electronic structure of MWSMs in a helical spin texture. We show that states are subjected to an effective sinusoidal potential in spin space, locally generating axial Landau levels below a certain energy threshold and find a condition for the generation of bulk flat bands. With this intuition, we are able to show that the nucleation of additional domain walls between contacts will modulate the conductance of the whole device, and that this effect persists for even trivial metallic electrodes.  Experimentally, such a system could potentially be realized in a MWSM helimagnet or via injection of chiral DWs into a magnetic racetrack \cite{parkin_chiralDW_2018} with mean-free-path larger than the superlattice length. 




\section{\label{sec:PW} Domain wall superlattices and plane-wave analysis}

We consider two models to elucidate the electronic structure and transport in periodically-magnetized MWSM superlattices: one using the continuum plane-wave basis model with a perfectly helical spin texture and another using the standard tight-binding approximation for transport. For both models, we consider the cases of N\'eel (N) and Bloch (B) magnetic DW lattices with the spatially-variant exchange field defined as 
\begin{align}
    \boldsymbol{\beta}_N(\boldsymbol{r}) =& \,J(\cos\theta_r\,\hat{y} + \sin\theta_r\,\hat{x}),\\
    \label{blochprofile}
    \boldsymbol{\beta}_B(\boldsymbol{r}) =& \,J(\cos\theta_r\,\hat{y} + \sin\theta_r\,\hat{z}),
\end{align}
where $\theta_r$ is a spatially-dependent magnetization angle. For the infinite superlattice along $\hat{x}$, $\theta_r$ can be defined as $\theta_r = G x$, where the reciprocal lattice length $G = 2\pi/a_s \text{, and } a_s = n_x a_0$ is the superlattice period of $n_x$ unit cells of length $a_0$. Later, we define $\theta_r$ for a series of chiral DWs. For the continuum model, we take the four-band model Hamiltonian of a TRS-broken MWSM~\cite{armitage_weyl_2018}:
\begin{equation}
\hat{H}(k)=\hbar v_f\tau_x \otimes (\mathbf{k}\cdot \boldsymbol{\sigma}) + m\tau_z \otimes \sigma_0 + \tau_0\otimes(\boldsymbol{\beta}\cdot\boldsymbol{\mathbf{\sigma}}),
\end{equation}
defined in the spin ($\sigma$) and pseudospin ($\tau$) spaces, where $v_f$ denotes the Fermi velocity and $m$ is a mass-like term. Imposing a 1D superlattice structure in $\hat{x}$ and adding the Fourier-transformed exchange field $\widetilde{\boldsymbol{\beta}}$ \footnote{See Supplemental Material at [URL will be inserted by publisher]} to the Hamiltonian off-diagonal terms result in

\begin{multline}
\label{eq:PWhamiltonian}
\hat{H}(k)=\sum_{G \in  \mathbb{Z} \vec{b}_x} c_{G}^{+}(\hbar v_f\tau_x \otimes \mathbf{(k-G)}\cdot \boldsymbol{\sigma} + m\tau_z \otimes \sigma_0)c_{G}   \\
+ \sum_{G', G \in \mathbb{Z}\vec{b}_x}c_{G'}^+(\tau_0 \otimes \widetilde{\boldsymbol{\beta}}_{(\mathbf{G}'-\mathbf{G})}\cdot\boldsymbol{\mathbf{\sigma}})c_{G}.
\end{multline}
In our model, we take $v_f = 1.5\times10^6 m/s$ \footnote{$v_f = 1.5 \times 10^6 m/s= \frac{a_0 t}{\hbar} = \frac{1 nm \times 1 eV}{\hbar}$. This is used for comparison with the tight-binding model. In real MWSMs, this is roughly an order of magnitude lower \cite{yamanouchi_srruo3_THT_2022}, but this does not change the underlying physics.} and focus on $m = 0$ in this paper. An exchange splitting magnitude of $J = 0.25 \; eV$ is taken to split the Weyl nodes with a small $\Delta k = 0.04 \frac{2\pi}{a_0}$ with $a_0 = 1$ nm, and for comparison with previous work~\cite{deSousa2021}. 

\begin{figure}[!ht]
\includegraphics[width=\linewidth]{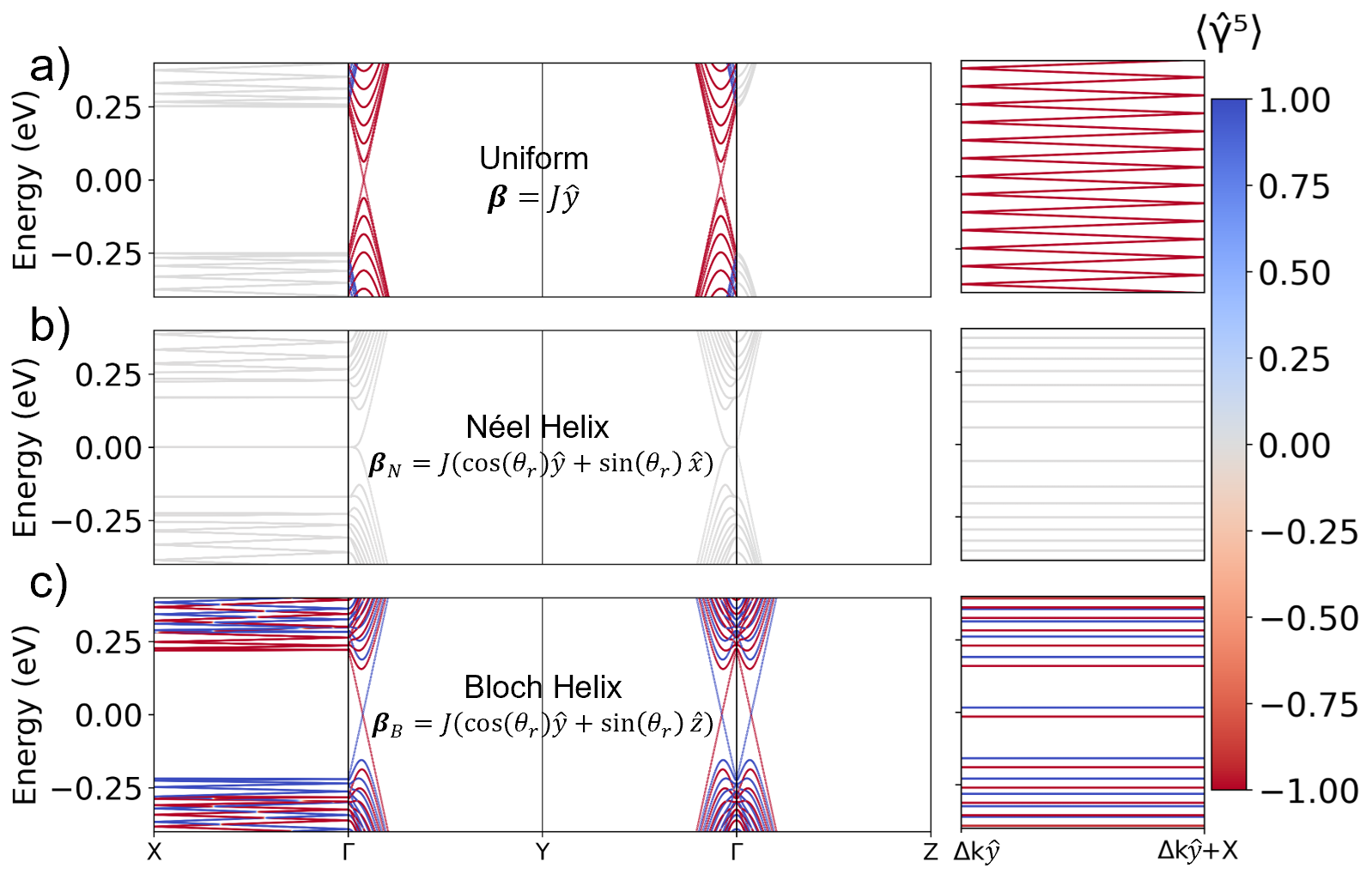}%
\caption{\label{fig:bandstructures} Chirality $\langle\gamma^5\rangle$ projected plane-wave band structures for a superlattice of period $a_s = 100$ nm and $ J = 0.25$ eV with differing exchange fields. $\Delta k \hat{y}$ denotes the position of the $|L\rangle$ Weyl point in the uniformly magnetized case, with the superlattice X point at $k=[\pi/a_S,0,0]$. (a) Uniformly magnetized MWSM shows $|L\rangle$ Weyl node with band folding. (b) N\'eel DW superlattices show band flattening in $\hat{x} \text{ and } \hat{y}$, while (c) Bloch DW superlattices only show band flattening in $\hat{x}$.}
\end{figure}


For a constant magnetization, the exchange field $\mathbf{\beta} = J \hat{M}$ will split the $|L\rangle \text{ or } |R\rangle$ chiral Weyl points (which we can denote with $\eta \in \{-1, +1\}$) in momentum space as $\mathbf{k}_\eta= -\eta \Delta k \hat{M} = - \eta\sqrt{J^2 - m^2} /\left(\hbar v_f\right)\hat{M}$. Then an equivalent low-energy model for the Weyl cone dispersions can be taken as \cite{armitage_weyl_2018,araki_localized_2018,grushin_inhomogeneous_weyl_2016, ilan_pseudo-electromagnetic_2020}

\begin{equation}
\hat{H}_{k,\eta} = \eta \hbar v_f [\mathbf{k\mathrm{+}k}_\eta] \cdot \boldsymbol{\sigma} = \eta \hbar v_f [\mathbf{k} - \eta \frac{e}{\hbar} \mathbf{A}^5]\cdot \boldsymbol{\sigma},
\end{equation}
%
where $\mathbf{A}^5$ leads to a chirality-dependent magnetic field determined by the exchange field texture:
\begin{equation}
\mathbf{B}^5 = \nabla \times \mathbf{A}^5 = \frac{J}{e v_f} \nabla \times \mathbf{M}(\mathbf{r}),
\end{equation}
which generates an axial magnetic field 
\begin{align}
\mathbf{B}^5_N(\boldsymbol{r}) =  \frac{-2\pi J}{a_{S} e v_f}(\sin\theta_r\hat{z})\\
\mathbf{B}^5_B(\boldsymbol{r}) = \frac{-2\pi J}{a_{S} e v_f} (\cos\theta_r\hat{y} + \sin\theta_r\hat{z})    
\end{align}
 for the N\'eel and Bloch wall cases, respectively. 
Applying this axial vector potential field to our system significantly modifies the band structure, generating flat bands and chirality-dependent axial Landau levels (LLs), shown in  Fig~\ref{fig:bandstructures}, where we plot the chirality-projected bands for the uniformly magnetized, N\'eel, and Bloch helical-exchange superlattices. The chirality operator $\gamma^5 = \tau_x\otimes\sigma_0$ has eigenvalues $\pm1$ corresponding to $\eta$, and $\langle \gamma^5 \rangle=\mathrm{Tr}(G^R_k(E)\gamma^5)/\mathrm{Tr}(G^R_k(E))$ is computed with the time-retarded Green's function $G^R_k = (E\hat{I}-\hat{H}_k)^{-1}$ in this section. While straightforward to project the Bloch eigenstates as $\langle u_n(k)|\gamma^5|u_n(k)\rangle $, it would be misleading in the N\'eel wall case as the $|L\rangle$ and $|R\rangle$ states are degenerate, thus obscuring one or the other. It is most interesting to note the generation of bulk flat bands in the direction of transport and axial LLs above the exchange-split momenta of the Weyl cones, which we will discuss further for each case. Both of these phenomena are explained in the low-energy model by the construction of Bloch functions in terms of localized harmonic oscillators created by the mixture of magnetic exchange and momentum terms. 

\subsection{Neel Helix Hamiltonian}

To understand the N\'eel Helix Hamiltonian $H_N$, it is first useful to gauge away the $\hat{x}$ term of $\mathbf{A}^5_N$, leaving the observables of the Hamiltonian unchanged: 
\begin{align}
\boldsymbol{\beta}'_N = J \text{cos}(Gx) \hat{y}; \\
\hat{H}_N = \hbar v_f \begin{bmatrix}
             \eta k_z & \Omega \\
            \Omega^+ & - \eta k_z\\
            \end{bmatrix}; \\
\Omega = -i\eta \partial_x - iJ\text{cos}(Gx) - i \eta k_y.
\end{align}
We may notice $\hat{H}_N^{2}$ is diagonal in spin space. Thus, we can motivate the wavefunctions $|\Psi_s\rangle$ for each spin $s\in{\pm 1}$.
\begin{align}
\hat{H}_N^2 = \left(\hbar v_f\right)^2 \begin{bmatrix}
             k_z^2 + \Omega\Omega^+ & 0 \\
            0 & k_z^2 + \Omega^+ \Omega \\
            \end{bmatrix}; \\
\left[ -\partial_x^2 + \frac{J\eta}{\hbar v_f} \left(2k_y\text{cos}(Gx)-sG\text{sin}(Gx)\right) +k_y^2 + \right. \nonumber \\ \left.k_z^2 + \left(\frac{J}{ \hbar v_f}\right)^2\frac{1+ \text{cos}(2 Gx)}{2}\right]|\Psi_s\rangle = \frac{E^2}{\left(\hbar v_f\right)^2}|\Psi_s\rangle.
\end{align}

\begin{figure}[]
\includegraphics[width=\linewidth]{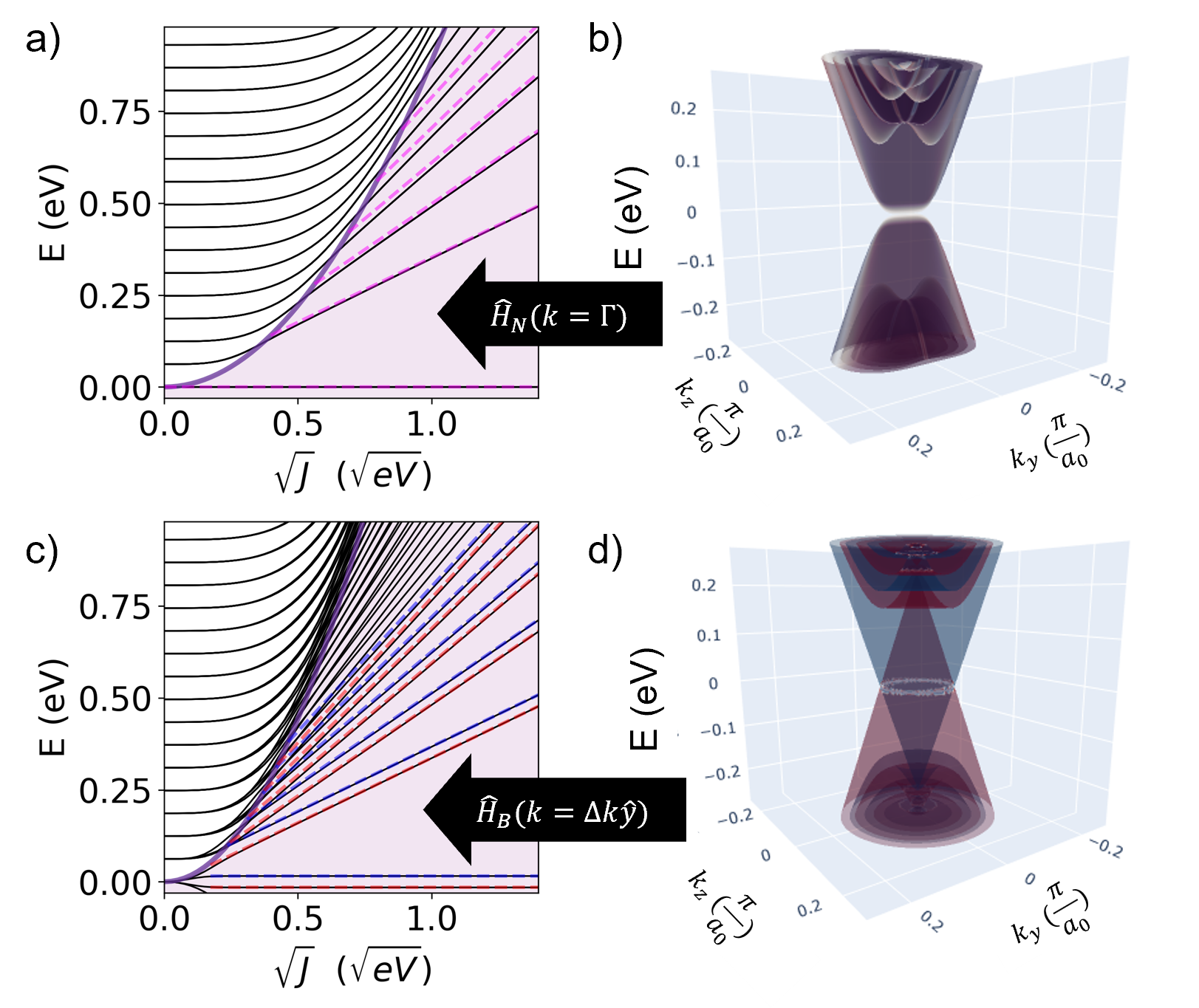}%
\caption{\label{fig:axialLLs} 
For a system with $a_S = 100$ nm, (a) and (c) show the spectrum of $\hat{H}_N(\Gamma)$ and $\hat{H}_B(\Delta k \hat{y})$, respectively, with the localization condition Eq.~\ref{bandflat} shown in purple. The analytic axial LL spacings are shown with dashed lines, in magenta, red, or blue for the degenerate, $|L\rangle$, or $|R\rangle$ bands. For a fixed $J=0.25$ eV, (b) and (d) show the corresponding 3D energy surfaces across the Brillouin zone at $k_x = 0$. 
}
\end{figure}

For realistic physical values of $J > \hbar v_f G$, and for small enough $k_y$, the above expressions reduce to the Mathieu differential equation from the dominant $\text{cos}(2 Gx)$ term -- notably without general analytic solutions. However, we can note that two locally-parabolic potential wells are created around regions of large $|\mathbf{B}^5|$, implying charge localization around $x = a_s(2\mathbb{Z}+1)/4$. Increasing small values of $k_y$ will tilt the potential wells, introducing anharmonicity away from $k = \Gamma$, but ultimately preserving the charge localization for small enough $k_y^2$. This is consistent with an intuition of a $\textbf{B}^5\propto \hat{z}$ which may localize charges in x and y. Around $k = \Gamma$, we can locally understand LL generation in $\hat{H}_N$ by defining the operators
\begin{align}
\hat{g} = \frac{1}{\sqrt{2 \hbar v_f J G}}\left(\eta v_f p_x - iJ\text{cos}(Gx)\right), \\
\hat{g}^+ = \frac{1}{\sqrt{2 \hbar v_f J G}}\left(\eta v_f p_x + iJ\text{cos}(Gx)\right),
\end{align}
which commute as 
\begin{align}
    \left[g,g^+\right] = -\eta \, \text{sin}(Gx) \approx \pm \left(1 - \text{O}(Gx)^2 + ...\right)
\end{align}
around the localized wavepackets. Thus, $g$ and $g^+$ switch roles as the creation and annihilation operators in different regions of the lattice and for differing electron chiralities. Linearizing the exchange term and mapping the problem onto the standard Dirac LL Hamiltonian, $\text{max}(\mathbf{B}^5)$ is predictive of the energy spacings for small enough n, as shown in Fig.~\ref{fig:axialLLs} (a):

\begin{align}
\text{max}(|\textbf{B}^5|) = \frac{G J}{e v_f}; \\
E_n \approx \text{sgn}(n) \sqrt{2|n| G J \hbar v_f + \left(\hbar v_f k_z\right)^2}. 
\label{neelLLspacing}
\end{align}

We may take as a guess that the 0th axial LL may have 0 energy at $k_{||} = \left[k_y, k_z\right] =\Gamma$, and proceed to directly solve for the null space of $\hat{H}'_{N}(\Gamma)$, leaving us with two expressions for the corresponding $|\Psi_\pm\rangle$ states.

\begin{eqnarray}
    \sqrt{2 \hbar v_f J G} \, \hat{g}|\Psi_-\rangle = 0 \nonumber \\
    \sqrt{2 \hbar v_f J G} \, \hat{g}^+|\Psi_+\rangle = 0 \nonumber. 
\end{eqnarray}

Finally, with a separation of variables, we can exactly solve for the zero-mode wavefunctions of each spin:

\begin{eqnarray}
    |\Psi_s\rangle = e^{\frac{\eta s J}{\hbar v_f G}\text{sin}(Gx)}|s\rangle.
\end{eqnarray}
which, around the potential wells, are well approximated by gaussians of characteristic magnetic length

\begin{equation}
    l_B = \sqrt{\frac{\hbar}{e|\mathbf{B}^5|}} = \sqrt{\frac{\hbar v_f}{G J}},
\end{equation}
thus acting as a well-defined 0th LL state. This tells us that the local axial LL model may hold in the limit where the local Bloch function spread (as determined by $l_B$) is much less than the superlattice length, which will be discussed shortly. 


\subsection{Bloch Helix Hamiltonian}

The prior analysis fails for the Bloch wall case, motivating us to consider other techniques to explain the behavior of $\hat{H}_B$. Qualitatively, we expect an analogous axial LL physics should manifest for the Bloch wall case, though the continuously changing direction of $\mathbf{B}^5$ makes it somewhat obscure: $\mathbf{B}^5$ rotates with constant magnitude in the $y-z$ plane as a function of $x$. As a result, the  zeroth Landau level disperses in different directions at well-defined positions in real-space, which can be seen in Fig.~\ref{fig:bandstructures} (c) for the $k_y$ and $k_z$ directions. We do observe that the magnetization profile $\boldsymbol{\beta}_B(\mathbf{r})$ (Eq.~\ref{blochprofile}), which decides the separation of two Weyl nodes, leads to a nodal-ring structure in the spectrum in the $k_y-k_z$ plane centered at $\Gamma$. To analyze this, first we will rotate $\hat{H}_B$ with $G$ oriented in $\hat{z}$, and translate the magnetic potential term while preserving its spatial helicity. Defining $k_x = k_{r} \text{cos}(\theta_k)$, $k_y = k_{r} \text{sin}(\theta_k)$ in cylindrical coordinates, and exploiting Euler's identity, we can write the transformed Hamiltonian as
\begin{align}
    \hat{H}_B = \begin{bmatrix}
             -i\eta \hbar v_f \partial_z  & \eta \hbar v_fk_{r} e^{-i\theta_k} + Je^{iGz} \\
            \eta \hbar v_f k_{r} e^{i\theta_k} + Je^{-iGz} &  i\eta \hbar v_f \partial_z\\
            \end{bmatrix}.
\end{align}
We can then change the basis to diagonalize the exchange term and square $H_B'$, giving us some insight on the form of the wavefunctions:

\begin{align} \label{chiralityfilter}
    \hat{H}_B^2 = -\eta \hbar v_f G J \hat{\sigma}_x - i G\hbar^2 v_f^2 \partial_z \left(\hat{\sigma}_0 - \hat{\sigma}_z\right) -\frac{\hbar^2 v_f^2 G^2}{4}\hat{\sigma}_z \nonumber \\
    +\left(2J^2 + 2\eta \hbar k_{r} v_f J\text{cos}(Gz+\theta_k)-\hbar^2v_f^2\partial_z^2+\frac{\hbar^2 v_f^2 G^2}{2}\right)\hat{\sigma}_0.
\end{align}

This implies that the Hamiltonian produces a continuum of locally-parabolic potential wells per unique $\theta_k$, and that the solutions will take the form of slightly modified Mathieu special functions for $J > \hbar v_f G$, this time with doubled spatial period and a single well per superlattice unit cell. Unfortunately, the off-diagonal term immensely complicates our analysis, mixing the solutions of the diagonal Mathieu ($s=1$) and Mathieu-like ($s=-1$) ODES. For further insight into the spin structure and of $\hat{H}_B$, we can replace $z$ with $x$ (for consistency with $\hat{H}_N$ analysis), define $\Lambda = \left(\eta k_{r} e^{i \theta_{k}} +Je^{-iGx} \right)^{-1}$ and solve for the components of $(\hat{H}_B-E)|\Psi\rangle = 0$ by substitution (see appendix \footnote{See Supplemental Material at [URL will be inserted by publisher] for full derivation.}). This is not analytically tractable in general, but we find two unnormalized solutions at $k_{r} = -\eta J / \left(\hbar v_f\right)$, $\theta_k = 0$, and $E = \eta\hbar v_f G/4$:

 \begin{align}
     |\Psi_\pm\rangle = \frac{1}{{\sqrt[4]{e^{i G x}}}} e^{\frac{\pm 2 J \left(1+e^{i G x}\right)}{G \hbar v_f \sqrt{e^{i G x}}}}|-\rangle 
     \mp \eta  \sqrt[4]{e^{i G x}} e^{\frac{\pm2 J \left(1+e^{i G x}\right)}{G \hbar v_f \sqrt{e^{i G x}}}}|+\rangle.
 \end{align}

We turn to numerics to find the spectrum of the system, and see that, as shown in Fig.~\ref{fig:axialLLs} (c),
\begin{equation}
    E_n \approx \text{sgn}(n) \sqrt{2|n| G J \hbar v_f} + \frac{\eta \hbar v_f G}{4}
\end{equation}
is indeed predictive at $|k_{r}| = J/\hbar v_f$. This implies the complex-modulated $E = \eta\hbar v_f G/4$ wavefunctions may indeed act as an exotic twisted 0th LL, however $H_B$ needs more analysis to determine a pair of operators which may obey the harmonic oscillator commutation relations. Interestingly, the dispersion for $|k_{||}| = k_r = 0$  can be obtained analytically \footnote{See Supplemental Material at [URL will be inserted by publisher] for full derivation} and is given by:
\begin{equation} \label{blochgammaenergies}
    E_{n}= \pm \sqrt{J^2 + \hbar^2 v_f^2 \left(k_x + \frac{\pi}{a_S}(2 n+1)\right)^2} + \frac{\eta \hbar v_f G}{2}.
\end{equation}
so that the lowest-energy $|R\rangle \text{ and }|L\rangle$-chiral bands are raised and lowered in energy with $\Delta E =\eta (\sqrt{J^2+\hbar^2 v_f^2 \pi^2/a_S^2} - \hbar v_f \pi/ a_S) \approx \eta J$ at the $\Gamma$ point for a magnetic texture with right-handed  spatial helicity and remain dispersive in $\hat{x}$. 


\subsection{Localization and band-flattening}

Both cases lead to a infinite lattice of locally-parabolic potential wells, which gives us a means to analyse the band flattening by considering  a minimal tight-binding model. For a localized LL state, 
\begin{equation}
    \langle \Delta x \rangle = l_B \sqrt{n+\frac{1}{2}},
\end{equation}
which provides a heuristic condition for band flattening as a function of axial LL index, supposing individual tightly-bound wavepackets may have negligible overlap. The Bloch and N\'eel wall cases provide 1 and 2 potential wells per unit cell respectively, thus: 

\begin{equation}
    \sqrt{\frac{\hbar v_f}{JG}\left(n+\frac{1}{2}\right)} << \begin{cases}
        \frac{a_S}{2} & \text{Neel Helix} \\ 
        a_S & \text{Bloch Helix} \\ 
    \end{cases}.
\end{equation}
Translating this into an energy cutoff, we expect to see the onset of carrier localization for 
\begin{align} \label{bandflat}
    |E_N| < \sqrt{J\left(\frac{\pi J}{4} - \frac{\hbar v_f}{a_S}\right)},
    |E_B| < \sqrt{J\left(\pi J - \frac{\hbar v_f}{a_S}\right)},
\end{align}
for the N\'eel and Bloch cases. As shown in Fig.~\ref{fig:axialLLs}, this approximation matches well with the calculated spectrum, predicting the onset of well-defined dirac LL energies.

In both cases, we understand that the electronic bands are significantly flattened in the direction of the superlattice by generating an emergent periodic potential for each spin, and provide a heuristic to predict the onset of the bulk flat band state. This shows that the application of spatially-varying or periodic magnetic textures in MWSMs provides a means to disable dispersion by effectively imposing a series of tunneling barriers in spin space, thus providing dynamic control over the electronic structure. With this in mind, we move to a device picture to assess one application of MWSM DW lattices.

\section{\label{sec:TB}Quantum transport and tight-binding analysis}

For transport and spatial resolution of the distorted Weyl cones, we employ a four-band, two Weyl point (see footnote \footnote{It is necessary to include a k-dependent mass term, as explained in \cite{kobayashi_helicity-protected_2018}, in order to create a single pair of Weyl cones near the center of the Brillouin zone at $\Delta k\hat{y}$ with $\Delta k = (1/a_0) \mathrm{cos}^{-1}(1-J^2/(2t^2))$\cite{deSousa2021}, though $m(k) \propto (1-\text{cos}(ka))$ is near-zero in the region of the Weyl points}) tight-binding model of a MWSM \cite{deSousa2021}, though instead  choose the $\tau_x$ and $\tau_z$ matrices for the momentum and mass terms, as in Eq.~\ref{eq:PWhamiltonian}:

\begin{multline}
    \hat{H}_{\mathrm{site \, i}} = \sum_{j\in\{x,y,z\}} [\frac{t}{2}(c_\text{i}^+ \hat{\tau}_z -  c_{\mathrm{i}+\hat{j}}^+ \hat{\tau}_z) \otimes \hat{\sigma}_0 c_\mathrm{i} -  \\ 
    \frac{-it}{2}c_{\mathrm{i}+\hat{j}}^+\hat{\tau}_x \otimes \hat{\sigma}_jc_\mathrm{i}
    +\text{h.c} ] + c^+_\mathrm{i}\tau_0\otimes(\boldsymbol{\beta}(r_\mathrm{i})\cdot\boldsymbol{\mathbf{\sigma}})c_\mathrm{i},
\end{multline}
with hopping parameter $t = $ 1 eV, lattice parameter $a_0 = $ 1 nm for simplicity, 100 total sites in x, and the Bloch phase prefactors applied to hoppings in $\hat{y}$ and $\hat{z}$ to construct a supercell $H(k_{||})$. Here, $\theta_r$ of a $180^{\circ}$ DW is defined in \cite{ohandley_modern_1999} and convolved with a semi-infinite Dirac comb to generate the magnetization angle (see Supp. Fig.~7):
\begin{equation}
\theta_r = (2\text{tan}^{-1}(\pi e^{-\frac{x+d}{d}})) \star \sum_{\mathrm{n}=0}^\infty \delta(\mathrm{n} a_S -x_0 - x),
\end{equation}
with $\star$ being the convolution operator, $x_0$ referring to the start position of the chiral DW lattice, the DW width $d = 8$ nm, and a truncated superlattice period $a_{s} = 30$ nm. While these parameters are not necessarily physical with relevance to the modelling of real magnetic racetrack systems, they are a minimal model that is computationally tractable and allows us to demonstrate underlying physics. In a real system, thicker domain walls or a longer $a_S$ would decrease max($\mathbf{B}^5$)~\cite{grushin_inhomogeneous_weyl_2016} thus decreasing the axial LL spacing and zone-folded subband spacing.

\begin{figure}[!h]
\includegraphics[width=\linewidth]{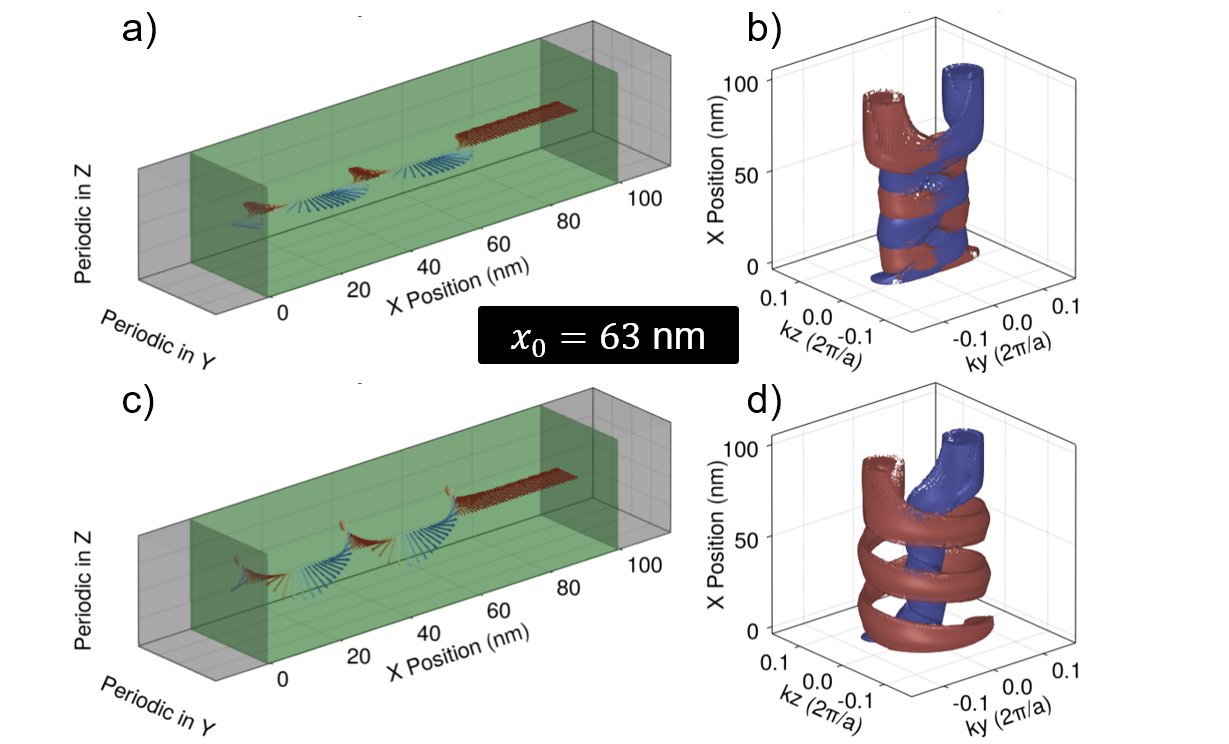}%
\caption{\label{fig:fermisurfaces} 
$\gamma^5$ and $k_{||}$ - resolved LDOS with a chemical potential $\mu = 0.1$ eV and broadening parameter $\eta = 10^{-2.5}$ eV are shown for the N\'eel (b) and Bloch (d) cases with corresponding exchange field textures $\boldsymbol{\beta}_N$ (a), $\boldsymbol{\beta}_B (c)$ for $x_0$ fixed to 63 nm. $|L\rangle$ - chiral states are shown in red, while $|R\rangle$-chiral states are shown in blue. 
}
\end{figure}


In the tight-binding model, we consider trivial metallic electrodes \footnote{We take a simple, spin and orbital-degenerate toy metal Hamiltonian $\hat{H}_{\mathrm{site}\;i} = \mathop{\sum_{d\in\{+,-\},j\in\{x,y,z\}}}{}\ [t c_{i+d\hat{j}}^+c_i \tau_0\otimes\sigma_0] + (3 t - \varepsilon_0) [c_i^+c_i\tau_0\otimes\sigma_0]$  with an arbitrary $t = 1$ eV and $\varepsilon_0 = 1$ eV} and semi-infinite MWSM electrodes to elucidate the underlying physics. For the case with MWSM electrodes, the magnetizations $\boldsymbol{\beta}(x=-a_0) \text{ and } \boldsymbol{\beta}(x=a_s+a_0)$ are copied for the left and right contact and extend to infinity for the generation of the Sancho-Rubio \cite{sancho_highly_1985} contact self-energies $\Sigma_{L,R}$ to retain the continuity of the magnetization field texture. 



\begin{figure}[!hb]
\includegraphics[width=\linewidth]{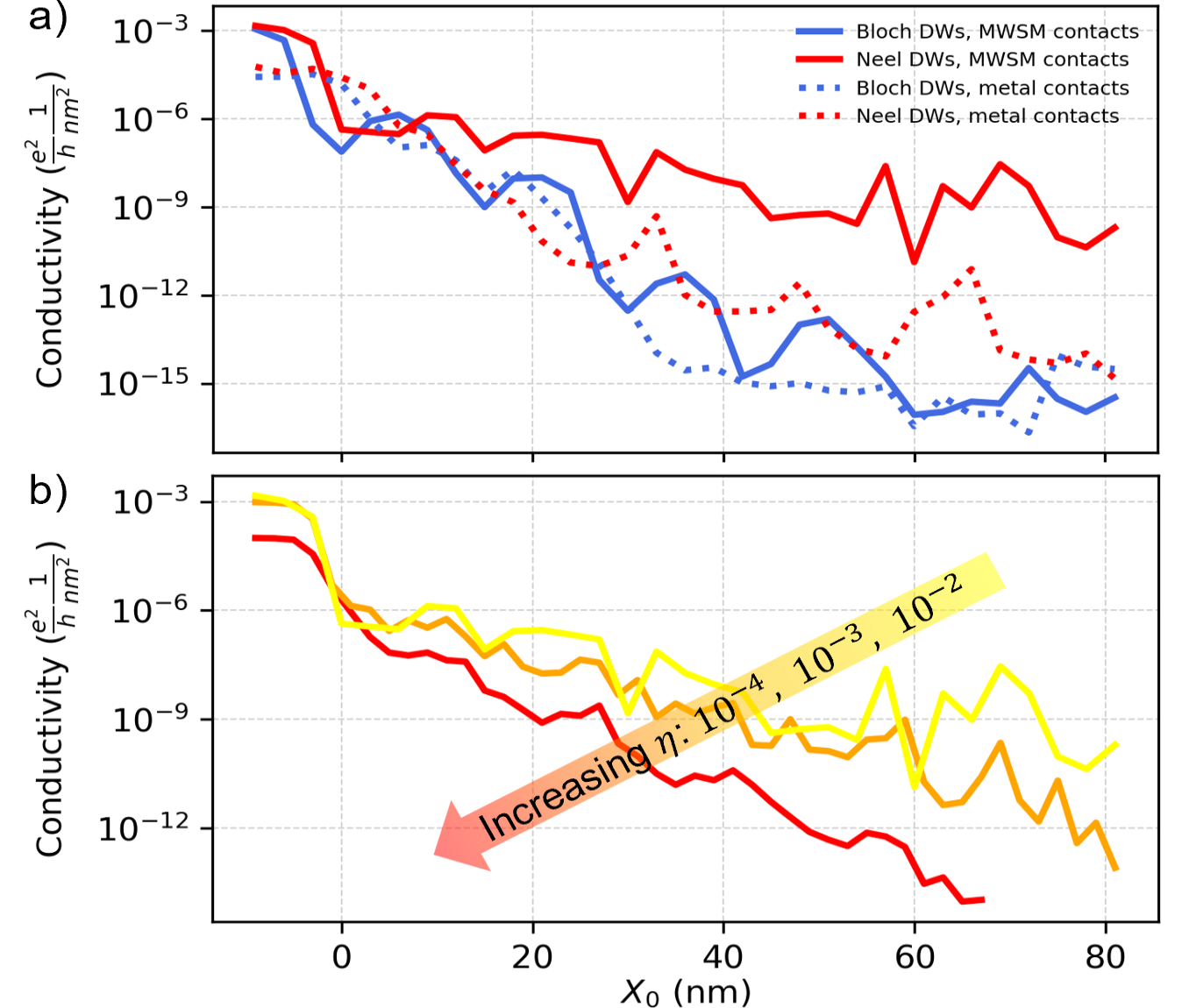}%
\caption{\label{fig:conductances}(a) Conductance modulation by orders of magnitude is shown for all combinations of MWSM and simple metallic electrodes with Bloch and N\'eel DW lattices by sweeping $x_0$. (b) A phenomenological broadening term $\eta$ is swept for the N\'eel wall case, showing discrete stairsteps in conductance, thus multi-weight behavior, as $\eta$ increases (i.e. carrier lifetime decreases).}
\end{figure}

\begin{figure*}[!ht]
\includegraphics[width=\linewidth]{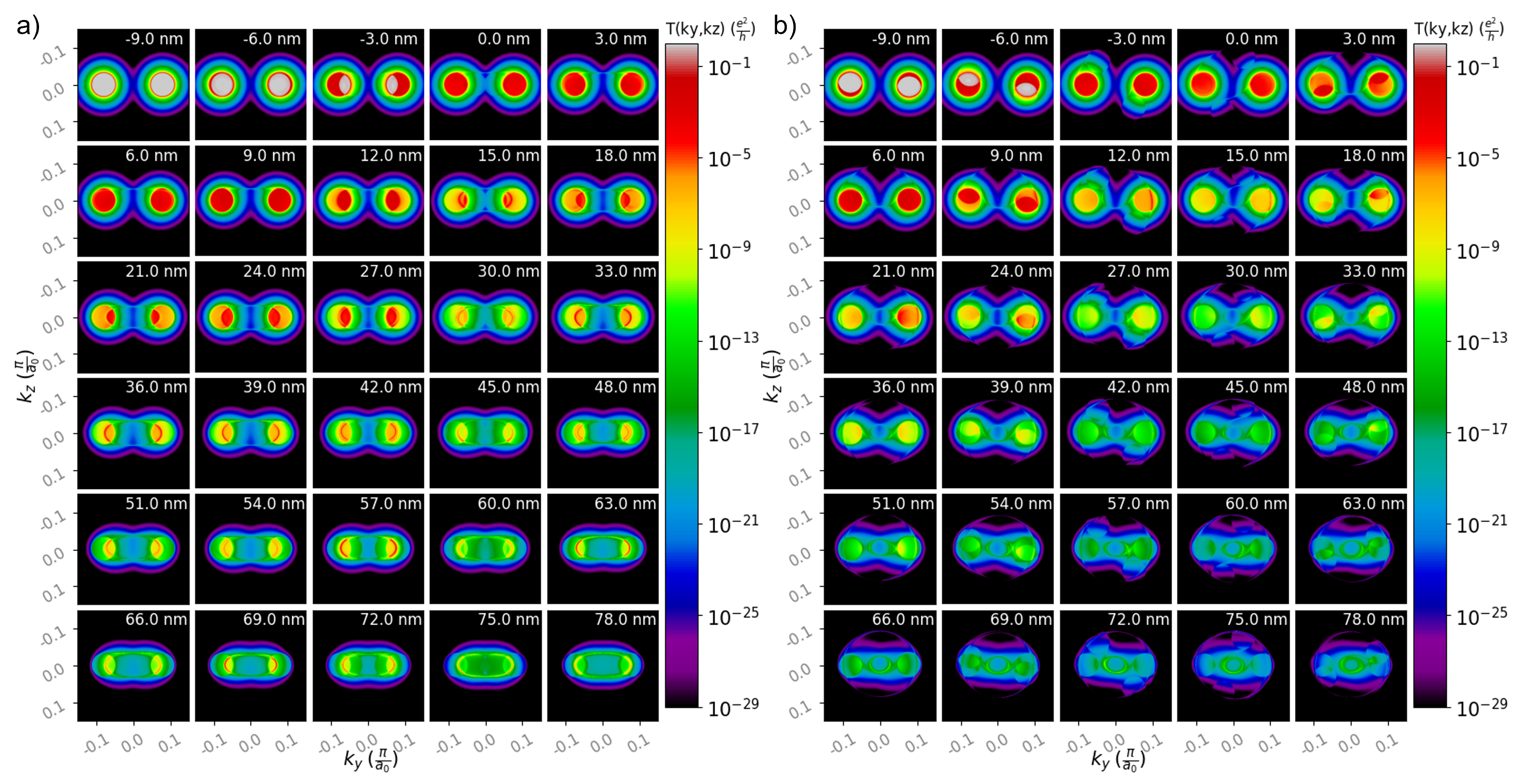}%
\caption{\label{fig:tmaps} (a) and (b) show k-resolved transmission maps for the N\'eel and Bloch DW cases, respectively, sweeping $x_0$, over the center of the surface Brillouin zone. Hotspots in transmission $T(k_{||})$ are visible where the bulk states from the left and right electrodes overlap, connected by twisted Fermi arcs of character determined by the magnetic texture. $x_0$ position is labeled in white for each sub-plot.}
\end{figure*}

 Of particular interest is the distortion of the Weyl cones in connection with the band structures of Fig.~\ref{fig:bandstructures}, which are useful to understand transport. For all transport calculations and tight-binding modelling, we take a chemical potential $\mu = 0.1$ eV to cut through the 0th axial LL from Sec. \ref{sec:PW}, and note that henceforth $\eta$ refers to the lifetime broadening. Fig.~\ref{fig:fermisurfaces} plots isosurfaces of the chirality, position, and $k_{||}$-resolved local density of states (LDOS) of the periodically magnetized racetrack devices, allowing us to spatially resolve the distorted Weyl cones which construct the bulk harmonic oscillator states. N\'eel devices show a $k_y$-symmetric twisting of the $|L\rangle$ and $|R\rangle$ Weyl cones, corresponding to the degenerate states in Fig.~\ref{fig:fermisurfaces} (b). This can be understood from the profile of $\boldsymbol{\beta}_N$ for the N\'eel case which decides the position of two Weyl nodes in the $k_y-k_z$ plane. On the other hand, for the Bloch DW case, $\mathbf{B}_B^5$ rotates in $y-z$, twisting the Weyl cones as shown in Fig.~\ref{fig:fermisurfaces} (d) where for an infinitesimal non-zero energy, $|L\rangle$ and $|R\rangle$ states are at a different radius in the $k_y-k_z$ plane. We see the $|L\rangle$ chiral states wrap around the expected position of the Weyl points at $\Delta k$, while the $|R\rangle$ chiral states wrap closely to $k_{||} = \Gamma$ for our chosen $\mu$, having been lifted up in energy from Eq.~\ref{blochgammaenergies}. These twisted Weyl cones motivate analysis of the chiral anomaly and Weyl orbits with periodic real and axial magnetic fields (see recent insights in \cite{breitkreiz_fermi_arc_2023}).

For device modeling, we neglect the orbital effects of the magnetic field and consider transport in the Landauer limit of the NEGF quantum transport formalism \footnote{See \cite{camsari_nonequilibrium_2023}; $G^R_{k_{||}} = (E+i\eta - H_{k_{||}} - \Sigma_{R,k_{||}} - \Sigma_{L,k_{||}})^{-1}; \Gamma_i = i(\Sigma_i - \Sigma_i^\dagger); T = tr(\Gamma_L G^R \Gamma_R G^A); G^A = (G^R)^\dagger$}. With a minimal carrier lifetime broadening of $\eta = 10^{-4}$ eV, in Fig.~\ref{fig:conductances} (a) we show that device conductance can be modulated by orders of magnitude via the injection of DWs into the active region of the device, and that this effect persists for both magnetization patterns, with both MWSM and simple metal electrodes. In Fig.~\ref{fig:conductances} (b), for a system with MWSM electrodes we sweep the broadening parameter $\eta$, corresponding to a phenomenological inelastic momentum-preserving scattering self-energy \cite{gilbert_TI_interconnects_2017}, and show that it 1) smooths out conductivity vs. $x_0$ and 2) forms stairsteps in conductance with respect to the number of DWs in the active region of the device--as one might expect with an increasing number of spin-dependent tunneling barriers. A similar behavior is observed for the Bloch wall case.

Curiously, even a high-resolution 401 x 401 k-grid over the zoomed-in portion of the surface Brillouin zone (SBZ) is unable to converge the value of the Landauer conductance for small broadening parameters ($\eta = 10^{-3}, 10^{-4}$ eV), leading to unphysical spikes in conduction. To explain this, we consider the $k_{||}$-resolved transmission with MWSM electrodes in Fig.~\ref{fig:tmaps} (a) and Fig.~\ref{fig:tmaps} (b) to resolve conductance behavior from the distorted Weyl cones in the periodic magnetization region. We see that the distorted Weyl cones in the bulk form infinitesimally-thick curtains of conductance in the SBZ which are challenging to capture numerically: In Fig.~\ref{fig:tmaps} (a), the N\'eel DW lattice conduction is dominated by the distorted $|L\rangle$ and $|R\rangle$ Weyl cones in the periodic region of the device, especially where they overlap with the contact electrodes' bulk Weyl cones. In contrast, the Bloch DW device in Fig.~\ref{fig:tmaps} (b) has minimal conduction through the distorted Weyl cones which wrap around the bulk Weyl cone states from either contact. Thus, the majority of the conductance tunnels through the periodic region of the device from the Bulk states in both contacts, leading to oscillations in Fig.~$\ref{fig:conductances}$ (a). The case with metal contacts in Fig.~$\ref{fig:conductances}$ (a) is somewhat more convoluted, given the formation of surface Fermi arcs at the metal/MWSM interface, though the properties of the device Hamiltonian and bulk Fermi arcs still dominate the qualitative behavior when sweeping $x_0$. Metal-contact $k_{||}$-resolved transmission plots are provided in the supplementary for both N\'eel and Bloch cases. Interestingly, due to the broadening parameter providing finite-lifetime states from both the MWSM and metal contacts, in our magnetization with left-handed spatial chirality, transport from the inner $|R\rangle$-chiral transmission ring will begin to dominate the total conductance, giving rise to a texture-dependent chirality filtering mechanism to generate or reflect chiral currents. In principle this would be tunable with a modulation in fermi level or magnetic texture, opening another means for generating chiral currents with potential in emerging devices~\cite{kharzeev_chiralanomaly_2013}. 


\section{\label{sec:discussion} Discussion}

We have explained the electronic structure and transport in a MWSM superlattice device, have shown magnetization-tunable axial LLs and flat bands in the direction of transport, and have demonstrated that multiple conductance states could potentially be encoded in such a device. We have also shown that the multi-state conduction behavior persists through all electrode configurations we have considered, unlike the spin transport in traditional spin filtering systems. With regards to the superlattice picture, the exploration of periodic exchange textures or artificial gauge fields and their topological structure is nascent in systems such as semiconductor nanowires \cite{kammhuber_helical_2017}, magnetic superlattice graphene \cite{wolf_periodicgraphene_2021}, or 2D Moir\'e magnets \cite{paul_skyrmionlattice_2023}. It should be noted that this effect only relies on the mean-field exchange splitting of a periodic ferromagnet, acting on spin space with contributions on the order of 100 to 1000 meV in bulk ferromagnets, as opposed to a real magnetic field or proximity-induced exchange, the effects of which would show with much weaker contribution to an effective gauge field. Thus, we expect the electronic structure in periodic MWSM lattices to be more robust closer to room temperatures, supposing a MWSM with Curie temperature above 300 K as in Mn\textsubscript{3}Sn\cite{muduli_mn3sn_2019}, Co\textsubscript{2}MnGa\cite{saito_co2mnga_2021}, or Fe\textsubscript{3}Sn\textsubscript{2}\cite{yao_fe3sn2_2018}. Nevertheless, other perspectives on scattering at MWSM DWs imply that skew scattering \cite{sorn_skewscattering_2021} or heavily-tilted Weyl cones \cite{xuan_tiltedMR_2021} could significantly decrease the MR. 

This highly-tunable electronic structure and conductivity via control of magnetization in MWSMs would be of great use to nanoelectronics, which we will discuss briefly. Even with additional achiral bands at the Fermi energy, non-magnetic electrodes, or a modest carrier relaxation length just greater than the width of the DW, a meager DW MR could be improved by chaining DWs along the length of the device to 1) improve the overall on/off ratio and 2) encode multiple bits of information in a single memory device. While it is hard to make definitive claims from our model without incoherent momentum relaxation -- we see longitudinal Weyl MRs over $10^{14} \%$ -- we estimate that, using the DW resistances observed in \cite{shiogai_co3sn2s2} an effective MR $= \left(R_{\text{off}}-R_{\text{on}}\right)/R_{\text{off}}=n_{DW}R_{DW}/R_0 = (\frac{L}{d})\frac{R_{DW}}{R_{0}}=  32.1\%$ could be achievable for the observed $R_{DW}\approx 4 \,\Omega$ and $R_{0}\approx 53 \,\Omega,$ for a domain width $d = 80$ nm and device length $L = 600$ nm. This is 14x the 3.93\% read margin observed in CoFeB transverse-read DW memory \cite{roxy_DWM_TR_2020}. In itself, this becomes competitive with other GMR devices (with regards to on/off ratio) or multi-state TMR devices \cite{leonard_synapses_2022}, but without the need of a magnetic tunnel junction to measure the magnetization-induced resistance change. Supposing a deliberate engineering of the chemical potential or DW width could bring about a modest DW MR $=25\%$ with $L = 1 \mu$m, and $d = 40$ nm, on/off ratios upwards of 650\% become possible. Another potential advantage of this approach, compared to traditional DW-MTJ devices, is that one could avoid the need for precise MgO deposition to avoid pinholes, thus making high MRs accessible to industry and academic labs without lengthy and expensive fabrication processes. Here, one could construct memory devices using simple metallic thin films, so long as the chirality-magnetization locking in the MWSM is preserved and surface fermi arcs do not dominate the conductance. Predicted orders-of-magnitude-improved readout performance over existing MTJs~\cite{deSousa2021} could further increase device viability, supposing one could amplify the Weyl MR or suppress scattering in a real system. 

\phantomsection{}
\textit{Note added: }Recently, we became aware of a paper which considers the Bloch helix case and describes this system as a 'Fermi Arc Metal', reporting further on the conductance in real magnetic fields \cite{breitkreiz_fermi_arc_2023}.

\section{Acknowledgements}
The authors acknowledge funding from the UT CDCM MRSEC supported under NSF Award Number DMR-1720595, funding from Sandia National Laboratories, and computational resources from the Texas Advanced Computing Center (TACC) at the University of Texas at Austin. The authors also greatly thank Eslam Khalaf and Gregory A. Fiete for helpful insights, as well as Kerem Camsari and Shuvro Chowdhury for their discussions on the theory and implementation of the NEGF transport formalism.

\bibliographystyle{apsrev4-1}
\bibliography{ref.bib}
\appendix
\include{supplementary4}
\end{document}

%% file: supplementary4.tex
\newcommand{\vket}[2]{
\begin{bmatrix} #1 \\ #2 \end{bmatrix}
}

\newcommand{\hbra}[2]{
\begin{bmatrix} #1  \; #2 \end{bmatrix}
}

\onecolumngrid
\appendix
\section{Supplementary information}

\subsection{Bloch wall hamiltonian spectrum with $k_{||} = 0$}
\label{suppblochgammaderivation}
First, one can choose the equivalent block-diagonal $\tau_z$ matrix for the momentum term in Eq.~\ref{eq:PWhamiltonian} and set m = 0. This decouples the left (-) and right (+) helical states $\eta$, and allows one to express the Hamiltonian of the infinitely-periodic hamiltonian with helical exchange field as follows. Let $\Omega = \frac{J}{2\hbar v_f}$, and let $G=\frac{2\pi}{a_S} = \vec{b}_x$ be the superlattice reciprocal vector. We use the definition for $\boldsymbol{\beta}(r)\boldsymbol{\beta}(r)$ in section II: 
\begin{eqnarray}
    \label{FT-exchange}
    \boldsymbol{\beta}_N(r) = J(\cos(\theta_r)\hat{y} + \sin(\theta_r)\hat{x}) = \frac{J}{2}(e^{i G x}\hat{y}+e^{-i G x}\hat{y} + -ie^{i G x}\hat{x}+ie^{-i G x}\hat{x}), \\\boldsymbol{\beta}_B(r) = J(\cos(\theta_r)\hat{y} + \sin(\theta_r)\hat{z}) = \frac{J}{2}(e^{i G x}\hat{y}+e^{-i G x}\hat{y} + -ie^{i G x}\hat{z}+ie^{-i G x}\hat{z}).
\end{eqnarray}

In Fourier space, this can be written as:
\begin{eqnarray}
    \widetilde{\boldsymbol{\beta}}^N_{G'-G} = \frac{J}{2}[(i\hat{x} + \hat{y})\delta(G'-G-\vec{b}_x) + (-i\hat{x} + \hat{y})\delta(G'-G+\vec{b}_x)]; \\
    \widetilde{\boldsymbol{\beta}}^B_{G'-G} = \frac{J}{2}[(i\hat{z} + \hat{y})\delta(G'-G-\vec{b}_x) + (-i\hat{z} + \hat{y})\delta(G'-G+\vec{b}_x)].
\end{eqnarray}

We will now focus on the Bloch wall case. 
\begin{equation}
    \hat{H}_B(k) = \hbar v_f\begin{bmatrix}
    \ddots & \Omega (\sigma_y + i \sigma_z) &    &      &    \\
     \Omega (\sigma_y - i \sigma_z) & \eta(k_x + G)\sigma_x       & \Omega (\sigma_y + i \sigma_z)   &     &    \\
      &         \Omega (\sigma_y - i \sigma_z)& \eta (k_x)\sigma_x         & \Omega (\sigma_y + i \sigma_z)    &     \\
      &         &           \Omega (\sigma_y - i \sigma_z)&     \eta (k_x - G)\sigma_x      & \Omega (\sigma_y + i \sigma_z)   \\
      &         &           &           \Omega (\sigma_y - i \sigma_z)& \ddots
  \end{bmatrix}
\end{equation}

Then, the basis transformation corresponding to a $120^{\circ}$ rotation about $\hat{x} +\hat{y} + \hat{z}$ can be used to transform the spin matrices:
\begin{eqnarray}
\sigma_y \rightarrow -\sigma_x  \nonumber \\
\sigma_z \rightarrow \sigma_y \nonumber \\
\sigma_x \rightarrow -\sigma_z \nonumber
\end{eqnarray}
and, knowing that $\sigma_+ = \sigma_x + i \sigma_y; \sigma_- = \sigma_x - i \sigma_y$, the Hamiltonian can be rewritten as:
\begin{equation}
    \hat{H}_B(k) = -\hbar v_f\begin{bmatrix}
    \ddots & 2\Omega (\sigma_-) &    &      &    \\
     2 \Omega (\sigma_+) & \eta(k_x + G)\sigma_z       & \Omega (\sigma_-)   &     &    \\
      &         2\Omega (\sigma_+)& \eta (k_x)\hat{\sigma}_z         & 2\Omega (\sigma_-)    &     \\
      &         &           2\Omega (\sigma_+)&     \eta (k_x - G)\hat{\sigma}_z      & 2\Omega (\sigma_-)   \\
      &         &           &           \Omega (\sigma_+)& \ddots
  \end{bmatrix}
\end{equation}
\begin{equation}
  = -\hbar v_f \bigoplus_n \left(\begin{bmatrix}
  -\eta(k_x + (n+1)G) & 2\Omega \\
  2\Omega & \eta(k_x + n G)
  \end{bmatrix}\right) = -\hbar v_f \bigoplus_n \left(-\eta k_x \sigma_z + 2 \Omega \sigma_x - \eta n G \sigma_z -\eta G\frac{1}{2}(\sigma_z+\sigma_0)\right)
\end{equation}

\begin{equation}
= -\hbar v_f \bigoplus_n (\icol{2\Omega \\ 0 \\ -\eta(k_x +  G (n + \frac{1}{2}))}\cdot\boldsymbol{\sigma} - \eta G \frac{1}{2}\sigma_0) = \bigoplus_n \hat{H}_n
\end{equation}
with the spectrum of $\hat{H}_n$ (thus, a diagonal subblock of $\hat{H}_B(k_{||}=\Gamma))$ given as:

\begin{equation}
    E_n = -\hbar v_f \left( \pm\sqrt{4\Omega^2 + (k_x + G(n+\frac{1}{2}))^2} - \eta G / 2\right) = \mp \sqrt{J^2 + \hbar^2 v_f^2 (k_x + \frac{\pi}{a_S}(2 n+1))^2} + \hbar v_f  \frac{\eta \pi}{a_S}
\end{equation}

\subsection{Decomposing Bloch exchange helix Hamiltonian}
First, it is useful to rotate the problem with $G=\frac{2\pi}{a_S}$ in $\hat{z}$ while preserving the Bloch magnetization $\boldsymbol{\beta}_B$'s helicity:

\begin{equation}
    \boldsymbol{\beta}_B = \text{J(cos}(G z)\hat{x} + \text{sin}(G z)\hat{y}
\end{equation}

Introducing $G^\pm = e^{\pm i G z}$ and absorbing $\hbar$ into $v_f$, we can rewrite the Hamiltonian $\hat{H}_B$ for one chirality $\eta$:

\begin{equation}
    \hat{H}_B = -i \eta v_f (\frac{\partial}{\partial x} \sigma_x + \frac{\partial}{\partial y} \sigma_y + \frac{\partial}{\partial z} \sigma_z) + J(G^+\sigma_+ + G^-\sigma_-)   
\end{equation}

We can then Fourier transform the Hamiltonian in x and y, and change the basis of $\hat{H}_B$ so as to diagonalize the helical exchange term:

\begin{align}
    \hat{H}'_B = \frac{1}{2}\begin{bmatrix}
        G^- & 1 \\
        -G^- & 1
    \end{bmatrix}
    \begin{bmatrix}
        -i \eta v_f \partial_z & \eta v_f k_x -i \eta v_f k_y + J G^+ \\
        \eta v_f k_x + i \eta v_f k_y + J G^- & i \eta v_f \partial_z \\
    \end{bmatrix} 
    \begin{bmatrix}
        G^+ & -G^+ \\
        1 & 1
    \end{bmatrix}. \\
    = \frac{1}{2}\begin{bmatrix}
        G^- & 1 \\
        -G^- & 1
    \end{bmatrix}
    \begin{bmatrix}
        A & B\\
        C & D \\
    \end{bmatrix}; \\
    A = G G^+\eta v_f - i \eta v_f G^+ \partial_z+JG^+ + \eta v_f \left(k_x-i k_y\right), \nonumber \\
    B = i \eta v_f G^+ \partial_z - G\eta v_f G^+ + JG^+ + \eta v_f \left(k_x - i k_y\right), \nonumber \\
    C = J + \eta v_f G^+\left(k_x +i k_y \right) + i \eta v_f \partial_z \nonumber, \\
    D = -J - \eta v_fG^+\left(k_x + i k_y\right) +i\eta v_f \partial_z;\nonumber \\ 
    \nonumber \\ 
    = \frac{1}{2}
    \begin{bmatrix}
        G^- A+C & G^-B+D\\
        -G^-A+C & -G^-B+D \\
    \end{bmatrix}; \\
    G^- A+C = G \eta v_f  + 2J + k_x \eta v_f \left(G^++G^-\right) + i k_y \eta v_f \left(G^+-G^-\right), \nonumber \\ 
    G^- B+D = 2 i \eta v_f \partial_z  - G \eta v_f  + k_x \eta v_f \left(-G^++G^-\right) - i k_y \eta v_f \left(G^++G^-\right), \nonumber \\ 
    -G^- A+C = - G \eta v_f  + 2 i \eta v_f \partial_z + k_x \eta v_f \left(G^+-G^-\right) + i k_y \eta v_f \left(G^++G^-\right), \nonumber \\ 
    -G^- B+D = G \eta v_f  - 2J - k_x \eta v_f \left(G^++G^-\right) - i k_y \eta v_f \left(G^+-G^-\right). \nonumber \\ 
\end{align}

Switching to cylindrical coordinates in x and y, we can define $k_x = \text{cos}(\theta_k)k_r$ and $k_y = \text{sin}(\theta_k)k_r$. We can also rewrite the above expressions: 

\begin{align}
    G^- A+C = G \eta v_f  + 2J + k_r \text{cos}(\theta_k) \eta v_f \left(G^++G^-\right) + i k_r \text{sin}(\theta_k \eta v_f \left(G^+-G^-\right), \nonumber \\ 
    G^- B+D = 2 i \eta v_f \partial_z  - G \eta v_f  + k_r \text{cos}(\theta_k) \eta v_f \left(-G^++G^-\right) - i k_r \text{sin}(\theta_k) \eta v_f \left(G^++G^-\right), \nonumber \\ 
    -G^- A+C = - G \eta v_f  + 2 i \eta v_f \partial_z + k_r \text{cos}(\theta_k) \eta v_f \left(G^+-G^-\right) + i k_r \text{sin}(\theta_k) \eta v_f \left(G^++G^-\right), \nonumber \\ 
    -G^- B+D = G \eta v_f  - 2J - k_r \text{cos}(\theta_k) \eta v_f \left(G^++G^-\right) - i k_r \text{sin}(\theta_k) \eta v_f \left(G^+-G^-\right); \nonumber \\ 
    \nonumber \\ 
    G^- A+C = G \eta v_f  + 2J + k_r \eta v_f \left( e^{i\theta_k} G^+ + e^{-i\theta_k}G^- \right), \nonumber \\ 
    G^- B+D = 2 i \eta v_f \partial_z  - G \eta v_f  +k_r \eta v_f \left(-e^{i\theta_k}G^+ + e^{-i\theta_k}G^-\right), \nonumber \\ 
    -G^- A+C = - G \eta v_f  + 2 i \eta v_f \partial_z + k_r \eta v_f \left(e^{i\theta_k}G^+ - e^{-i\theta_k}G^-\right), \nonumber \\ 
    -G^- B+D = G \eta v_f  - 2J - k_r \eta v_f \left( e^{i\theta_k} G^+ + e^{-i\theta_k}G^- \right); \nonumber \\ 
    \nonumber \\
    \hat{H}'_B=
    \begin{bmatrix}
        \frac{G \eta v_f}{2} + J + k_r \eta v_f \text{cos}(Gz + \theta_k) &  i \eta v_f \partial_z  - \frac{G \eta v_f}{2}  -i k_r \eta v_f \text{sin}(Gz + \theta_k)\\
        i \eta v_f \partial_z  - \frac{G \eta v_f}{2} + i k_r \eta v_f \text{sin}(Gz + \theta_k) & \frac{G \eta v_f}{2} - J - k_r \eta v_f \text{cos}(Gz + \theta_k) \\
    \end{bmatrix} \nonumber \\
    = \frac{G \eta v_f}{2}\hat{\sigma}_0 + \left(J+k_r\eta v_f \text{cos}(Gz + \theta_k)\right)\hat{\sigma}_z + \left(i \eta v_f \partial_z  - \frac{G \eta v_f}{2}\right)\hat{\sigma}_x +k_r\eta v_f \text{sin}(Gz + \theta_k)\hat{\sigma}_y
\end{align}

We can notice that $\hat{H}_B$ permits a continuous set of spatially translated solutions for each value of $\theta_k$. Here, we will also stop and note the correspondence with \ref{suppblochgammaderivation} for $k_r = 0$. We may also notice that a $k_r$ corresponding to the position of the uniformly-magnetized $|L\rangle$-chiral weyl node,

\begin{equation}
    k_r = \frac{-\eta J}{v_f},
\end{equation}
will cancel the spatially constant terms of $\hat{\sigma}_z$ using a taylor series around $\theta_k=0$ in z:
\begin{align}
    \hat{H}'_B\left(k=\frac{-\eta J}{v_f}\right)  = \frac{G \eta v_f}{2}\hat{\sigma}_0 + J\frac{\left(Gz\right)^2}{2}\hat{\sigma}_z + \left(i \eta v_f \partial_z  - \frac{G \eta v_f}{2}\right)\hat{\sigma}_x - J G z \hat{\sigma}_y.
\end{align}
Giving us the insight that the Hamiltonian is locally akin to the Dirac Hamiltonian with an additional off-diagonal term which, if the wavefunctions are represented in a hermite-gauss function basis, will generate $|n\pm2\rangle$ states. Squaring $H'_B$ and again changing the basis ($\sigma_x \rightarrow \sigma_z$, $\sigma_z \rightarrow -\sigma_x$) allows us to decouple the spatially-varying terms of the Hamiltonian for each spin, giving us some insight as explained in the main text:

\begin{align}
    H_B^{'2} = \left(\frac{\eta \hbar v_f G}{2}\right)^2 + 2\left(\frac{\eta \hbar v_f G}{2}\right)\Omega + \Omega^2 \nonumber \\  
    = \frac{\hbar^2 v_f^2 G^2}{4}\hat{\sigma}_0+\hbar^2 v_f^2 G k_r\left[\text{sin}(Gz+\theta_k)\hat{\sigma}_y +  \text{cos}(Gz+\theta_k)\hat{\sigma}_z\right]-\hbar^2 v_f^2 G k_r\left[\text{sin}(Gz+\theta_k)\hat{\sigma}_y +  \text{cos}(Gz+\theta_k)\hat{\sigma}_z\right] \nonumber \\ +\left[\frac{\hbar^2 v_f^2 G^2}{4} -\hbar^2 v_f^2 \left(\partial_z^2 +i G \partial_z\right) + 2J^2+2\hbar^2 k_r^2\text{cos}(Gz+\theta_k)\right]\hat{\sigma}_0+i\hbar^2 v_f^2 G \partial_z \hat{\sigma}_x + \eta \hbar v_f G J\hat{\sigma}_z -\frac{\hbar^2 v_f^2 G^2}{4}\hat{\sigma}_x.\\
    \rightarrow \nonumber \\ 
    \hat{H}_B^2 = -\eta \hbar v_f G J \hat{\sigma}_x - i G\hbar^2 v_f^2 \partial_z \left(\hat{\sigma}_0 - \hat{\sigma}_z\right) -\frac{\hbar^2 v_f^2 G^2}{4}\hat{\sigma}_z +\left(2J^2 + 2\eta k_r v_f \text{cos}(Gz+\theta_k)-\hbar^2v_f^2\partial_z^2+\frac{\hbar^2 v_f^2 G^2}{2}\right)\hat{\sigma}_0.
\end{align}

We are unable to solve this exactly, but changing $z \rightarrow x$, can manipulate the spin terms of $\left(H_B-E\right)|\Psi\rangle = 0$ to generate constraints on the wavefunctions, defining $\Lambda = \left(\eta \hbar v_f k_r e^{i \theta_{k}} +Je^{-iGx} \right)^{-1}$.
\begin{align}
    \left[\left(\hbar v_f\right)^2 \left(\Lambda^+ \partial_x^2 + \frac{\partial \Lambda^+}{\partial x}\partial_x \right) +i\eta \hbar v_f \frac{\partial \Lambda^+}{\partial x}E - \Lambda^+ E^2+ \frac{1}{\Lambda}\right]|\uparrow\rangle = 0, \\ 
    \left[\left(\hbar v_f\right)^2 \left(\Lambda \partial_x^2 + \frac{\partial \Lambda}{\partial x}\partial_x \right) -i\eta \hbar v_f \frac{\partial \Lambda}{\partial x}E - \Lambda E^2+ \frac{1}{\Lambda^+}\right]|\downarrow\rangle  = 0,\label{downbloch}\\
    |\downarrow\rangle = -\Lambda^+ \left(-i\eta v_f \partial_x - E\right)|\uparrow\rangle,\\
    |\uparrow\rangle = -\Lambda \left(i\eta v_f \partial_x - E\right)|\downarrow\rangle. \label{upbloch}
\end{align}

 Composing Eq. \ref{downbloch} and Eq. \ref{upbloch} at $k_{r} = \frac{-\eta J}{\hbar v_f}$ and $E = \eta \hbar v_f G/4$, we can use the Wolfram Mathematica software to find two solutions for a $|\Psi\rangle$ with normalization coefficients $\mathbb{C}_1, \mathbb{C}_2$:

 \begin{align}
     |\Psi\rangle = \frac{1}{{\sqrt[4]{e^{i G x}}}}\left[\mathbb{C}_1 e^{\frac{2 J \left(1+e^{i G x}\right)}{G \hbar v_f \sqrt{e^{i G x}}}}+\mathbb{C}_2 e^{-\frac{2 J \left(1+e^{i G x}\right)}{G \hbar v_f \sqrt{e^{i G x}}}}\right]|-\rangle \nonumber \\
     -\eta  \sqrt[4]{e^{i G x}}  \left[\mathbb{C}_1 e^{\frac{2 J \left(1+e^{i G x}\right)}{G \hbar v_f \sqrt{e^{i G x}}}}-\mathbb{C}_2 e^{-\frac{2 J \left(1+e^{i G x}\right)}{G \hbar v_f \sqrt{e^{i G x}}}}\right]|+\rangle.
 \end{align}

 \subsection{Tight-binding transport supplementary figures}

\begin{figure}[!ht]
\includegraphics[width=0.5\columnwidth]{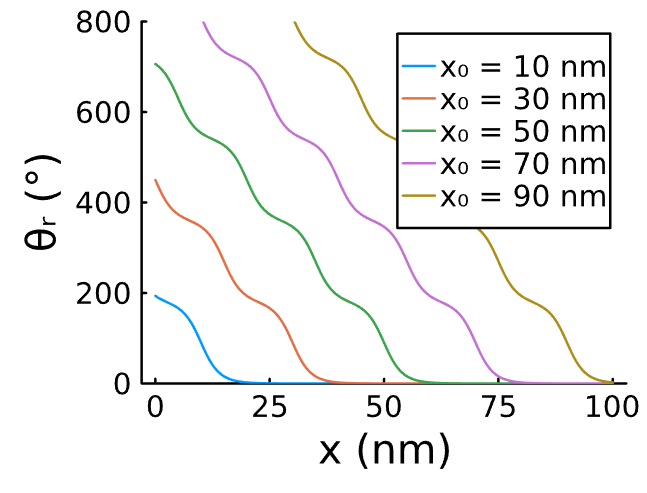}%
\caption{\label{fig:supp_theta_r} The magnetization angle $\theta_r$ of the tight-binding Hamiltonian is shown above for varying values of $x_0$, where $x_0$ intuits the start position of a series of domain walls.}
\end{figure}

\begin{figure*}[!ht]
\includegraphics[width=0.5\columnwidth]{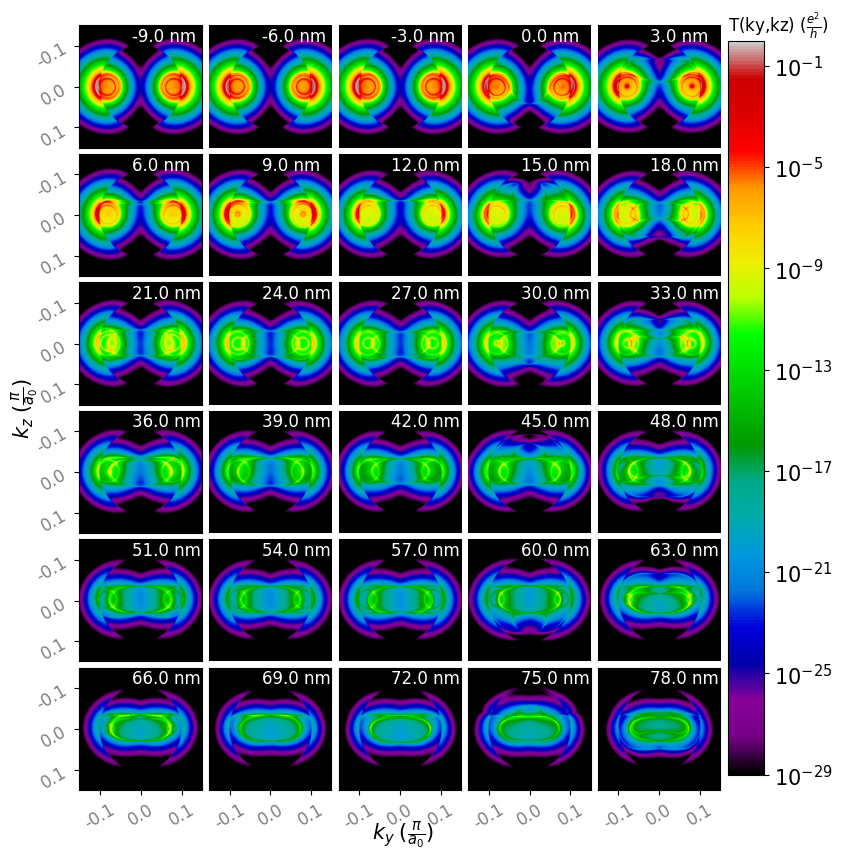}%
\caption{\label{fig:tmaps_metal_neel} Shown above are the $k_{||}$-resolved transmission maps for the Ne\'el wall case with metallic electrodes, sweeping $x_0$. One can see additional states injected from the electrodes, which hybridize with the Fermi arcs inside the MWSM}
\end{figure*}

\begin{figure*}[!ht]
\includegraphics[width=0.5\columnwidth]{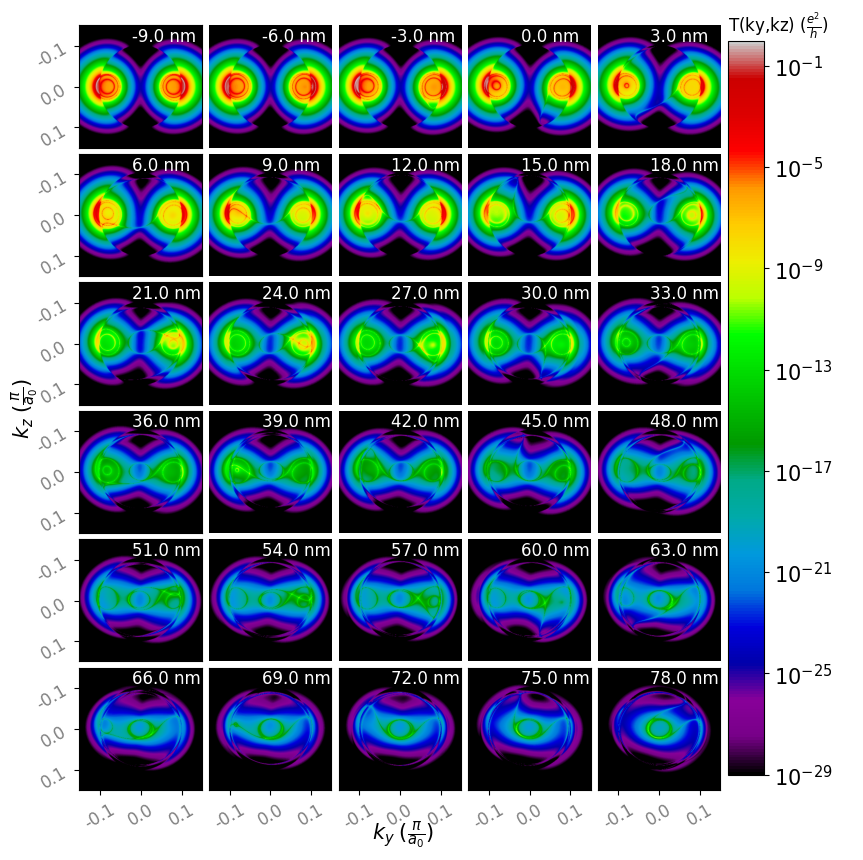}%
\caption{\label{fig:tmaps_metal_bloch} $k_{||}$-resolved transmission maps are shown for the Bloch wall case with metallic electrodes, sweeping $x_0$.}
\end{figure*}

%% file: draftv0.bbl
\begin{thebibliography}{48}%
\makeatletter
\providecommand \@ifxundefined [1]{%
 \@ifx{#1\undefined}
}%
\providecommand \@ifnum [1]{%
 \ifnum #1\expandafter \@firstoftwo
 \else \expandafter \@secondoftwo
 \fi
}%
\providecommand \@ifx [1]{%
 \ifx #1\expandafter \@firstoftwo
 \else \expandafter \@secondoftwo
 \fi
}%
\providecommand \natexlab [1]{#1}%
\providecommand \enquote  [1]{``#1''}%
\providecommand \bibnamefont  [1]{#1}%
\providecommand \bibfnamefont [1]{#1}%
\providecommand \citenamefont [1]{#1}%
\providecommand \href@noop [0]{\@secondoftwo}%
\providecommand \href [0]{\begingroup \@sanitize@url \@href}%
\providecommand \@href[1]{\@@startlink{#1}\@@href}%
\providecommand \@@href[1]{\endgroup#1\@@endlink}%
\providecommand \@sanitize@url [0]{\catcode `\\12\catcode `\$12\catcode
  `\&12\catcode `\#12\catcode `\^12\catcode `\_12\catcode `\%12\relax}%
\providecommand \@@startlink[1]{}%
\providecommand \@@endlink[0]{}%
\providecommand \url  [0]{\begingroup\@sanitize@url \@url }%
\providecommand \@url [1]{\endgroup\@href {#1}{\urlprefix }}%
\providecommand \urlprefix  [0]{URL }%
\providecommand \Eprint [0]{\href }%
\providecommand \doibase [0]{http://dx.doi.org/}%
\providecommand \selectlanguage [0]{\@gobble}%
\providecommand \bibinfo  [0]{\@secondoftwo}%
\providecommand \bibfield  [0]{\@secondoftwo}%
\providecommand \translation [1]{[#1]}%
\providecommand \BibitemOpen [0]{}%
\providecommand \bibitemStop [0]{}%
\providecommand \bibitemNoStop [0]{.\EOS\space}%
\providecommand \EOS [0]{\spacefactor3000\relax}%
\providecommand \BibitemShut  [1]{\csname bibitem#1\endcsname}%
\let\auto@bib@innerbib\@empty
\bibitem [{\citenamefont {Armitage}\ \emph {et~al.}(2018)\citenamefont
  {Armitage}, \citenamefont {Mele},\ and\ \citenamefont
  {Vishwanath}}]{armitage_weyl_2018}%
  \BibitemOpen
  \bibfield  {author} {\bibinfo {author} {\bibfnamefont {N.}~\bibnamefont
  {Armitage}}, \bibinfo {author} {\bibfnamefont {E.}~\bibnamefont {Mele}}, \
  and\ \bibinfo {author} {\bibfnamefont {A.}~\bibnamefont {Vishwanath}},\
  }\href {\doibase 10.1103/RevModPhys.90.015001} {\bibfield  {journal}
  {\bibinfo  {journal} {Reviews of Modern Physics}\ }\textbf {\bibinfo {volume}
  {90}},\ \bibinfo {pages} {015001} (\bibinfo {year} {2018})}\BibitemShut
  {NoStop}%
\bibitem [{\citenamefont {Burkov}(2018)}]{burkov2018weyl}%
  \BibitemOpen
  \bibfield  {author} {\bibinfo {author} {\bibfnamefont {A.}~\bibnamefont
  {Burkov}},\ }\href
  {https://www.annualreviews.org/doi/abs/10.1146/annurev-conmatphys-033117-054129?casa_token=5lpJgYxWS2cAAAAA\%3AjN-dy0JC2T20ETWZnnIcGIKJYZasvnKHa-yMJw25qKK_UVmv2HHFS4CY3pl7MsiEmymxm7MvZI_qRA}
  {\bibfield  {journal} {\bibinfo  {journal} {Annual Review of Condensed Matter
  Physics}\ }\textbf {\bibinfo {volume} {9}},\ \bibinfo {pages} {359} (\bibinfo
  {year} {2018})}\BibitemShut {NoStop}%
\bibitem [{\citenamefont {Ilan}\ \emph {et~al.}(2020)\citenamefont {Ilan},
  \citenamefont {Grushin},\ and\ \citenamefont
  {Pikulin}}]{ilan_pseudo-electromagnetic_2020}%
  \BibitemOpen
  \bibfield  {author} {\bibinfo {author} {\bibfnamefont {R.}~\bibnamefont
  {Ilan}}, \bibinfo {author} {\bibfnamefont {A.~G.}\ \bibnamefont {Grushin}}, \
  and\ \bibinfo {author} {\bibfnamefont {D.~I.}\ \bibnamefont {Pikulin}},\
  }\href {\doibase 10.1038/s42254-019-0121-8} {\bibfield  {journal} {\bibinfo
  {journal} {Nature Reviews Physics}\ }\textbf {\bibinfo {volume} {2}},\
  \bibinfo {pages} {29} (\bibinfo {year} {2020})}\BibitemShut {NoStop}%
\bibitem [{\citenamefont {Hu}\ \emph {et~al.}(2019)\citenamefont {Hu},
  \citenamefont {Xu}, \citenamefont {Ni},\ and\ \citenamefont
  {Mao}}]{hu2019transport}%
  \BibitemOpen
  \bibfield  {author} {\bibinfo {author} {\bibfnamefont {J.}~\bibnamefont
  {Hu}}, \bibinfo {author} {\bibfnamefont {S.-Y.}\ \bibnamefont {Xu}}, \bibinfo
  {author} {\bibfnamefont {N.}~\bibnamefont {Ni}}, \ and\ \bibinfo {author}
  {\bibfnamefont {Z.}~\bibnamefont {Mao}},\ }\href
  {https://www.annualreviews.org/doi/abs/10.1146/annurev-matsci-070218-010023?casa_token=8hl_D3SbS40AAAAA\%3AAA4sfm27pHzpiESruEA8Wcnl-2gBhwHKrqbSVe9fDytKlSCS3yfZ2fTux5J7cxAAYg4knOWRRIqx}
  {\bibfield  {journal} {\bibinfo  {journal} {Annual Review of Materials
  Research}\ }\textbf {\bibinfo {volume} {49}},\ \bibinfo {pages} {207}
  (\bibinfo {year} {2019})}\BibitemShut {NoStop}%
\bibitem [{\citenamefont {Nagaosa}\ \emph {et~al.}(2020)\citenamefont
  {Nagaosa}, \citenamefont {Morimoto},\ and\ \citenamefont
  {Tokura}}]{nagaosa2020transport}%
  \BibitemOpen
  \bibfield  {author} {\bibinfo {author} {\bibfnamefont {N.}~\bibnamefont
  {Nagaosa}}, \bibinfo {author} {\bibfnamefont {T.}~\bibnamefont {Morimoto}}, \
  and\ \bibinfo {author} {\bibfnamefont {Y.}~\bibnamefont {Tokura}},\ }\href
  {https://www.nature.com/articles/s41578-020-0208-y} {\bibfield  {journal}
  {\bibinfo  {journal} {Nature Reviews Materials}\ }\textbf {\bibinfo {volume}
  {5}},\ \bibinfo {pages} {621} (\bibinfo {year} {2020})}\BibitemShut {NoStop}%
\bibitem [{\citenamefont {de~Sousa}\ \emph {et~al.}(2021)\citenamefont
  {de~Sousa}, \citenamefont {Ascencio}, \citenamefont {Haney}, \citenamefont
  {Wang},\ and\ \citenamefont {Low}}]{deSousa2021}%
  \BibitemOpen
  \bibfield  {author} {\bibinfo {author} {\bibfnamefont {D.~J.~P.}\
  \bibnamefont {de~Sousa}}, \bibinfo {author} {\bibfnamefont {C.~O.}\
  \bibnamefont {Ascencio}}, \bibinfo {author} {\bibfnamefont {P.~M.}\
  \bibnamefont {Haney}}, \bibinfo {author} {\bibfnamefont {J.~P.}\ \bibnamefont
  {Wang}}, \ and\ \bibinfo {author} {\bibfnamefont {T.}~\bibnamefont {Low}},\
  }\href {\doibase 10.1103/PhysRevB.104.L041401} {\bibfield  {journal}
  {\bibinfo  {journal} {Phys. Rev. B}\ }\textbf {\bibinfo {volume} {104}},\
  \bibinfo {pages} {L041401} (\bibinfo {year} {2021})}\BibitemShut {NoStop}%
\bibitem [{\citenamefont {Kobayashi}\ \emph {et~al.}(2018)\citenamefont
  {Kobayashi}, \citenamefont {Ominato},\ and\ \citenamefont
  {Nomura}}]{kobayashi_helicity-protected_2018}%
  \BibitemOpen
  \bibfield  {author} {\bibinfo {author} {\bibfnamefont {K.}~\bibnamefont
  {Kobayashi}}, \bibinfo {author} {\bibfnamefont {Y.}~\bibnamefont {Ominato}},
  \ and\ \bibinfo {author} {\bibfnamefont {K.}~\bibnamefont {Nomura}},\ }\href
  {\doibase 10.7566/JPSJ.87.073707} {\bibfield  {journal} {\bibinfo  {journal}
  {Journal of the Physical Society of Japan}\ }\textbf {\bibinfo {volume}
  {87}},\ \bibinfo {pages} {073707} (\bibinfo {year} {2018})}\BibitemShut
  {NoStop}%
\bibitem [{\citenamefont {Kent}\ \emph {et~al.}(2001)\citenamefont {Kent},
  \citenamefont {Yu}, \citenamefont {Rüdiger},\ and\ \citenamefont
  {Parkin}}]{kent_DWR_2001}%
  \BibitemOpen
  \bibfield  {author} {\bibinfo {author} {\bibfnamefont {A.~D.}\ \bibnamefont
  {Kent}}, \bibinfo {author} {\bibfnamefont {J.}~\bibnamefont {Yu}}, \bibinfo
  {author} {\bibfnamefont {U.}~\bibnamefont {Rüdiger}}, \ and\ \bibinfo
  {author} {\bibfnamefont {S.~S.~P.}\ \bibnamefont {Parkin}},\ }\href {\doibase
  10.1088/0953-8984/13/25/202} {\bibfield  {journal} {\bibinfo  {journal}
  {Journal of Physics: Condensed Matter}\ }\textbf {\bibinfo {volume} {13}},\
  \bibinfo {pages} {R461} (\bibinfo {year} {2001})}\BibitemShut {NoStop}%
\bibitem [{\citenamefont {Roxy}\ \emph {et~al.}(2020)\citenamefont {Roxy},
  \citenamefont {Ollivier}, \citenamefont {Hoque}, \citenamefont {Longofono},
  \citenamefont {Jones},\ and\ \citenamefont {Bhanja}}]{roxy_DWM_TR_2020}%
  \BibitemOpen
  \bibfield  {author} {\bibinfo {author} {\bibfnamefont {K.}~\bibnamefont
  {Roxy}}, \bibinfo {author} {\bibfnamefont {S.}~\bibnamefont {Ollivier}},
  \bibinfo {author} {\bibfnamefont {A.}~\bibnamefont {Hoque}}, \bibinfo
  {author} {\bibfnamefont {S.}~\bibnamefont {Longofono}}, \bibinfo {author}
  {\bibfnamefont {A.~K.}\ \bibnamefont {Jones}}, \ and\ \bibinfo {author}
  {\bibfnamefont {S.}~\bibnamefont {Bhanja}},\ }\href {\doibase
  10.1109/TNANO.2020.3014091} {\bibfield  {journal} {\bibinfo  {journal}
  {{IEEE} Transactions on Nanotechnology}\ }\textbf {\bibinfo {volume} {19}},\
  \bibinfo {pages} {648} (\bibinfo {year} {2020})}\BibitemShut {NoStop}%
\bibitem [{\citenamefont {Chen}\ \emph {et~al.}(2021)\citenamefont {Chen},
  \citenamefont {Li}, \citenamefont {Li},\ and\ \citenamefont
  {Miao}}]{chen_multiply_2021}%
  \BibitemOpen
  \bibfield  {author} {\bibinfo {author} {\bibfnamefont {J.}~\bibnamefont
  {Chen}}, \bibinfo {author} {\bibfnamefont {J.}~\bibnamefont {Li}}, \bibinfo
  {author} {\bibfnamefont {Y.}~\bibnamefont {Li}}, \ and\ \bibinfo {author}
  {\bibfnamefont {X.}~\bibnamefont {Miao}},\ }\href {\doibase
  10.1088/1674-4926/42/1/013104} {\bibfield  {journal} {\bibinfo  {journal}
  {Journal of Semiconductors}\ }\textbf {\bibinfo {volume} {42}},\ \bibinfo
  {pages} {013104} (\bibinfo {year} {2021})}\BibitemShut {NoStop}%
\bibitem [{\citenamefont {Li}\ \emph {et~al.}(2018)\citenamefont {Li},
  \citenamefont {Wang}, \citenamefont {Midya}, \citenamefont {Xia},\ and\
  \citenamefont {Yang}}]{li_memristor_review_2018}%
  \BibitemOpen
  \bibfield  {author} {\bibinfo {author} {\bibfnamefont {Y.}~\bibnamefont
  {Li}}, \bibinfo {author} {\bibfnamefont {Z.}~\bibnamefont {Wang}}, \bibinfo
  {author} {\bibfnamefont {R.}~\bibnamefont {Midya}}, \bibinfo {author}
  {\bibfnamefont {Q.}~\bibnamefont {Xia}}, \ and\ \bibinfo {author}
  {\bibfnamefont {J.~J.}\ \bibnamefont {Yang}},\ }\href {\doibase
  10.1088/1361-6463/aade3f} {\bibfield  {journal} {\bibinfo  {journal} {Journal
  of Physics D: Applied Physics}\ }\textbf {\bibinfo {volume} {51}},\ \bibinfo
  {pages} {503002} (\bibinfo {year} {2018})}\BibitemShut {NoStop}%
\bibitem [{\citenamefont {Wan}\ \emph {et~al.}(2022)\citenamefont {Wan},
  \citenamefont {Kubendran}, \citenamefont {Schaefer}, \citenamefont
  {Eryilmaz}, \citenamefont {Zhang}, \citenamefont {Wu}, \citenamefont {Deiss},
  \citenamefont {Raina}, \citenamefont {Qian}, \citenamefont {Gao},
  \citenamefont {Joshi}, \citenamefont {Wu}, \citenamefont {Wong},\ and\
  \citenamefont {Cauwenberghs}}]{wan_compute_memory_2022}%
  \BibitemOpen
  \bibfield  {author} {\bibinfo {author} {\bibfnamefont {W.}~\bibnamefont
  {Wan}}, \bibinfo {author} {\bibfnamefont {R.}~\bibnamefont {Kubendran}},
  \bibinfo {author} {\bibfnamefont {C.}~\bibnamefont {Schaefer}}, \bibinfo
  {author} {\bibfnamefont {S.~B.}\ \bibnamefont {Eryilmaz}}, \bibinfo {author}
  {\bibfnamefont {W.}~\bibnamefont {Zhang}}, \bibinfo {author} {\bibfnamefont
  {D.}~\bibnamefont {Wu}}, \bibinfo {author} {\bibfnamefont {S.}~\bibnamefont
  {Deiss}}, \bibinfo {author} {\bibfnamefont {P.}~\bibnamefont {Raina}},
  \bibinfo {author} {\bibfnamefont {H.}~\bibnamefont {Qian}}, \bibinfo {author}
  {\bibfnamefont {B.}~\bibnamefont {Gao}}, \bibinfo {author} {\bibfnamefont
  {S.}~\bibnamefont {Joshi}}, \bibinfo {author} {\bibfnamefont
  {H.}~\bibnamefont {Wu}}, \bibinfo {author} {\bibfnamefont {H.-S.~P.}\
  \bibnamefont {Wong}}, \ and\ \bibinfo {author} {\bibfnamefont
  {G.}~\bibnamefont {Cauwenberghs}},\ }\href {\doibase
  10.1038/s41586-022-04992-8} {\bibfield  {journal} {\bibinfo  {journal}
  {Nature}\ }\textbf {\bibinfo {volume} {608}},\ \bibinfo {pages} {504}
  (\bibinfo {year} {2022})}\BibitemShut {NoStop}%
\bibitem [{\citenamefont {Wang}\ \emph
  {et~al.}(2022{\natexlab{a}})\citenamefont {Wang}, \citenamefont {Si},
  \citenamefont {Jiang}, \citenamefont {Malik}, \citenamefont {Pan},
  \citenamefont {Stathopoulos}, \citenamefont {Serb}, \citenamefont {Wang},
  \citenamefont {Prodromakis},\ and\ \citenamefont
  {Papavassiliou}}]{wang_multistate_memristors_2022}%
  \BibitemOpen
  \bibfield  {author} {\bibinfo {author} {\bibfnamefont {C.}~\bibnamefont
  {Wang}}, \bibinfo {author} {\bibfnamefont {Z.}~\bibnamefont {Si}}, \bibinfo
  {author} {\bibfnamefont {X.}~\bibnamefont {Jiang}}, \bibinfo {author}
  {\bibfnamefont {A.}~\bibnamefont {Malik}}, \bibinfo {author} {\bibfnamefont
  {Y.}~\bibnamefont {Pan}}, \bibinfo {author} {\bibfnamefont {S.}~\bibnamefont
  {Stathopoulos}}, \bibinfo {author} {\bibfnamefont {A.}~\bibnamefont {Serb}},
  \bibinfo {author} {\bibfnamefont {S.}~\bibnamefont {Wang}}, \bibinfo {author}
  {\bibfnamefont {T.}~\bibnamefont {Prodromakis}}, \ and\ \bibinfo {author}
  {\bibfnamefont {C.}~\bibnamefont {Papavassiliou}},\ }\href {\doibase
  10.1109/JETCAS.2022.3223295} {\bibfield  {journal} {\bibinfo  {journal}
  {{IEEE} Journal on Emerging and Selected Topics in Circuits and Systems}\
  }\textbf {\bibinfo {volume} {12}},\ \bibinfo {pages} {723} (\bibinfo {year}
  {2022}{\natexlab{a}})}\BibitemShut {NoStop}%
\bibitem [{\citenamefont {Zarcone}\ \emph {et~al.}(2020)\citenamefont
  {Zarcone}, \citenamefont {Engel}, \citenamefont {Burc~Eryilmaz},
  \citenamefont {Wan}, \citenamefont {Kim}, \citenamefont {{BrightSky}},
  \citenamefont {Lam}, \citenamefont {Lung}, \citenamefont {Olshausen},\ and\
  \citenamefont {Philip~Wong}}]{zarcone_analog_2020}%
  \BibitemOpen
  \bibfield  {author} {\bibinfo {author} {\bibfnamefont {R.~V.}\ \bibnamefont
  {Zarcone}}, \bibinfo {author} {\bibfnamefont {J.~H.}\ \bibnamefont {Engel}},
  \bibinfo {author} {\bibfnamefont {S.}~\bibnamefont {Burc~Eryilmaz}}, \bibinfo
  {author} {\bibfnamefont {W.}~\bibnamefont {Wan}}, \bibinfo {author}
  {\bibfnamefont {S.}~\bibnamefont {Kim}}, \bibinfo {author} {\bibfnamefont
  {M.}~\bibnamefont {{BrightSky}}}, \bibinfo {author} {\bibfnamefont
  {C.}~\bibnamefont {Lam}}, \bibinfo {author} {\bibfnamefont {H.-L.}\
  \bibnamefont {Lung}}, \bibinfo {author} {\bibfnamefont {B.~A.}\ \bibnamefont
  {Olshausen}}, \ and\ \bibinfo {author} {\bibfnamefont {H.-S.}\ \bibnamefont
  {Philip~Wong}},\ }\href {\doibase 10.1038/s41598-020-63723-z} {\bibfield
  {journal} {\bibinfo  {journal} {Scientific Reports}\ }\textbf {\bibinfo
  {volume} {10}},\ \bibinfo {pages} {6831} (\bibinfo {year}
  {2020})}\BibitemShut {NoStop}%
\bibitem [{\citenamefont {Jung}\ \emph {et~al.}(2022)\citenamefont {Jung},
  \citenamefont {Lee}, \citenamefont {Myung}, \citenamefont {Kim},
  \citenamefont {Yoon}, \citenamefont {Kwon}, \citenamefont {Ju}, \citenamefont
  {Kim}, \citenamefont {Yi}, \citenamefont {Han}, \citenamefont {Kwon},
  \citenamefont {Seo}, \citenamefont {Lee}, \citenamefont {Koh}, \citenamefont
  {Lee}, \citenamefont {Song}, \citenamefont {Choi}, \citenamefont {Ham},\ and\
  \citenamefont {Kim}}]{jung_crossbar_2022}%
  \BibitemOpen
  \bibfield  {author} {\bibinfo {author} {\bibfnamefont {S.}~\bibnamefont
  {Jung}}, \bibinfo {author} {\bibfnamefont {H.}~\bibnamefont {Lee}}, \bibinfo
  {author} {\bibfnamefont {S.}~\bibnamefont {Myung}}, \bibinfo {author}
  {\bibfnamefont {H.}~\bibnamefont {Kim}}, \bibinfo {author} {\bibfnamefont
  {S.~K.}\ \bibnamefont {Yoon}}, \bibinfo {author} {\bibfnamefont {S.-W.}\
  \bibnamefont {Kwon}}, \bibinfo {author} {\bibfnamefont {Y.}~\bibnamefont
  {Ju}}, \bibinfo {author} {\bibfnamefont {M.}~\bibnamefont {Kim}}, \bibinfo
  {author} {\bibfnamefont {W.}~\bibnamefont {Yi}}, \bibinfo {author}
  {\bibfnamefont {S.}~\bibnamefont {Han}}, \bibinfo {author} {\bibfnamefont
  {B.}~\bibnamefont {Kwon}}, \bibinfo {author} {\bibfnamefont {B.}~\bibnamefont
  {Seo}}, \bibinfo {author} {\bibfnamefont {K.}~\bibnamefont {Lee}}, \bibinfo
  {author} {\bibfnamefont {G.-H.}\ \bibnamefont {Koh}}, \bibinfo {author}
  {\bibfnamefont {K.}~\bibnamefont {Lee}}, \bibinfo {author} {\bibfnamefont
  {Y.}~\bibnamefont {Song}}, \bibinfo {author} {\bibfnamefont {C.}~\bibnamefont
  {Choi}}, \bibinfo {author} {\bibfnamefont {D.}~\bibnamefont {Ham}}, \ and\
  \bibinfo {author} {\bibfnamefont {S.~J.}\ \bibnamefont {Kim}},\ }\href
  {\doibase 10.1038/s41586-021-04196-6} {\bibfield  {journal} {\bibinfo
  {journal} {Nature}\ }\textbf {\bibinfo {volume} {601}},\ \bibinfo {pages}
  {211} (\bibinfo {year} {2022})}\BibitemShut {NoStop}%
\bibitem [{\citenamefont {Brigner}\ \emph {et~al.}(2022)\citenamefont
  {Brigner}, \citenamefont {Hassan}, \citenamefont {Hu}, \citenamefont
  {Bennett}, \citenamefont {Garcia-Sanchez}, \citenamefont {Marinella},
  \citenamefont {Incorvia},\ and\ \citenamefont
  {Friedman}}]{brigner_purelyspintronic_2022}%
  \BibitemOpen
  \bibfield  {author} {\bibinfo {author} {\bibfnamefont {W.~H.}\ \bibnamefont
  {Brigner}}, \bibinfo {author} {\bibfnamefont {N.}~\bibnamefont {Hassan}},
  \bibinfo {author} {\bibfnamefont {X.}~\bibnamefont {Hu}}, \bibinfo {author}
  {\bibfnamefont {C.~H.}\ \bibnamefont {Bennett}}, \bibinfo {author}
  {\bibfnamefont {F.}~\bibnamefont {Garcia-Sanchez}}, \bibinfo {author}
  {\bibfnamefont {M.~J.}\ \bibnamefont {Marinella}}, \bibinfo {author}
  {\bibfnamefont {J.~A.~C.}\ \bibnamefont {Incorvia}}, \ and\ \bibinfo {author}
  {\bibfnamefont {J.~S.}\ \bibnamefont {Friedman}},\ }in\ \href {\doibase
  10.1109/ISCAS48785.2022.9937890} {\emph {\bibinfo {booktitle} {2022 {IEEE}
  International Symposium on Circuits and Systems ({ISCAS})}}}\ (\bibinfo
  {year} {2022})\ pp.\ \bibinfo {pages} {1189--1193}\BibitemShut {NoStop}%
\bibitem [{\citenamefont {Shiogai}\ \emph {et~al.}(2022)\citenamefont
  {Shiogai}, \citenamefont {Ikeda}, \citenamefont {Fujiwara}, \citenamefont
  {Seki}, \citenamefont {Takanashi},\ and\ \citenamefont
  {Tsukazaki}}]{shiogai_co3sn2s2}%
  \BibitemOpen
  \bibfield  {author} {\bibinfo {author} {\bibfnamefont {J.}~\bibnamefont
  {Shiogai}}, \bibinfo {author} {\bibfnamefont {J.}~\bibnamefont {Ikeda}},
  \bibinfo {author} {\bibfnamefont {K.}~\bibnamefont {Fujiwara}}, \bibinfo
  {author} {\bibfnamefont {T.}~\bibnamefont {Seki}}, \bibinfo {author}
  {\bibfnamefont {K.}~\bibnamefont {Takanashi}}, \ and\ \bibinfo {author}
  {\bibfnamefont {A.}~\bibnamefont {Tsukazaki}},\ }\href {\doibase
  10.1103/PhysRevMaterials.6.114203} {\bibfield  {journal} {\bibinfo  {journal}
  {Phys. Rev. Mater.}\ }\textbf {\bibinfo {volume} {6}},\ \bibinfo {pages}
  {114203} (\bibinfo {year} {2022})}\BibitemShut {NoStop}%
\bibitem [{\citenamefont {Araki}\ \emph {et~al.}(2018)\citenamefont {Araki},
  \citenamefont {Yoshida},\ and\ \citenamefont
  {Nomura}}]{araki_localized_2018}%
  \BibitemOpen
  \bibfield  {author} {\bibinfo {author} {\bibfnamefont {Y.}~\bibnamefont
  {Araki}}, \bibinfo {author} {\bibfnamefont {A.}~\bibnamefont {Yoshida}}, \
  and\ \bibinfo {author} {\bibfnamefont {K.}~\bibnamefont {Nomura}},\ }\href
  {\doibase 10.1103/PhysRevB.98.045302} {\bibfield  {journal} {\bibinfo
  {journal} {Physical Review B}\ }\textbf {\bibinfo {volume} {98}},\ \bibinfo
  {pages} {045302} (\bibinfo {year} {2018})}\BibitemShut {NoStop}%
\bibitem [{\citenamefont {Grushin}\ \emph {et~al.}(2016)\citenamefont
  {Grushin}, \citenamefont {Venderbos}, \citenamefont {Vishwanath},\ and\
  \citenamefont {Ilan}}]{grushin_inhomogeneous_weyl_2016}%
  \BibitemOpen
  \bibfield  {author} {\bibinfo {author} {\bibfnamefont {A.~G.}\ \bibnamefont
  {Grushin}}, \bibinfo {author} {\bibfnamefont {J.~W.~F.}\ \bibnamefont
  {Venderbos}}, \bibinfo {author} {\bibfnamefont {A.}~\bibnamefont
  {Vishwanath}}, \ and\ \bibinfo {author} {\bibfnamefont {R.}~\bibnamefont
  {Ilan}},\ }\href {\doibase 10.1103/PhysRevX.6.041046} {\bibfield  {journal}
  {\bibinfo  {journal} {Phys. Rev. X}\ }\textbf {\bibinfo {volume} {6}},\
  \bibinfo {pages} {041046} (\bibinfo {year} {2016})}\BibitemShut {NoStop}%
\bibitem [{\citenamefont {Kurebayashi}\ and\ \citenamefont
  {Nomura}(2019)}]{kurebayashi_theory_2019}%
  \BibitemOpen
  \bibfield  {author} {\bibinfo {author} {\bibfnamefont {D.}~\bibnamefont
  {Kurebayashi}}\ and\ \bibinfo {author} {\bibfnamefont {K.}~\bibnamefont
  {Nomura}},\ }\href {\doibase 10.1038/s41598-019-41776-z} {\bibfield
  {journal} {\bibinfo  {journal} {Scientific Reports}\ }\textbf {\bibinfo
  {volume} {9}},\ \bibinfo {pages} {5365} (\bibinfo {year} {2019})}\BibitemShut
  {NoStop}%
\bibitem [{\citenamefont {Yamanouchi}\ \emph {et~al.}(2022)\citenamefont
  {Yamanouchi}, \citenamefont {Araki}, \citenamefont {Sakai}, \citenamefont
  {Uemura}, \citenamefont {Ohta},\ and\ \citenamefont
  {Ieda}}]{yamanouchi_srruo3_THT_2022}%
  \BibitemOpen
  \bibfield  {author} {\bibinfo {author} {\bibfnamefont {M.}~\bibnamefont
  {Yamanouchi}}, \bibinfo {author} {\bibfnamefont {Y.}~\bibnamefont {Araki}},
  \bibinfo {author} {\bibfnamefont {T.}~\bibnamefont {Sakai}}, \bibinfo
  {author} {\bibfnamefont {T.}~\bibnamefont {Uemura}}, \bibinfo {author}
  {\bibfnamefont {H.}~\bibnamefont {Ohta}}, \ and\ \bibinfo {author}
  {\bibfnamefont {J.}~\bibnamefont {Ieda}},\ }\href {\doibase
  10.1126/sciadv.abl6192} {\bibfield  {journal} {\bibinfo  {journal} {Science
  Advances}\ }\textbf {\bibinfo {volume} {8}},\ \bibinfo {pages} {eabl6192}
  (\bibinfo {year} {2022})}\BibitemShut {NoStop}%
\bibitem [{\citenamefont {Wang}\ \emph
  {et~al.}(2022{\natexlab{b}})\citenamefont {Wang}, \citenamefont {Zeng},
  \citenamefont {Yuan}, \citenamefont {Zeng}, \citenamefont {Gu}, \citenamefont
  {Xu}, \citenamefont {Wang}, \citenamefont {Han}, \citenamefont {Nomura},
  \citenamefont {Wang}, \citenamefont {Liu}, \citenamefont {Hou},\ and\
  \citenamefont {Ye}}]{wang_co3sn2s2_THT_2022}%
  \BibitemOpen
  \bibfield  {author} {\bibinfo {author} {\bibfnamefont {Q.}~\bibnamefont
  {Wang}}, \bibinfo {author} {\bibfnamefont {Y.}~\bibnamefont {Zeng}}, \bibinfo
  {author} {\bibfnamefont {K.}~\bibnamefont {Yuan}}, \bibinfo {author}
  {\bibfnamefont {Q.}~\bibnamefont {Zeng}}, \bibinfo {author} {\bibfnamefont
  {P.}~\bibnamefont {Gu}}, \bibinfo {author} {\bibfnamefont {X.}~\bibnamefont
  {Xu}}, \bibinfo {author} {\bibfnamefont {H.}~\bibnamefont {Wang}}, \bibinfo
  {author} {\bibfnamefont {Z.}~\bibnamefont {Han}}, \bibinfo {author}
  {\bibfnamefont {K.}~\bibnamefont {Nomura}}, \bibinfo {author} {\bibfnamefont
  {W.}~\bibnamefont {Wang}}, \bibinfo {author} {\bibfnamefont {E.}~\bibnamefont
  {Liu}}, \bibinfo {author} {\bibfnamefont {Y.}~\bibnamefont {Hou}}, \ and\
  \bibinfo {author} {\bibfnamefont {Y.}~\bibnamefont {Ye}},\ }\href {\doibase
  10.1038/s41928-022-00879-8} {\bibfield  {journal} {\bibinfo  {journal}
  {Nature Electronics}\ ,\ \bibinfo {pages} {1}} (\bibinfo {year}
  {2022}{\natexlab{b}})}\BibitemShut {NoStop}%
\bibitem [{\citenamefont {Bechler}\ and\ \citenamefont
  {Masell}()}]{bechler_helitronics_2023}%
  \BibitemOpen
  \bibfield  {author} {\bibinfo {author} {\bibfnamefont {N.~T.}\ \bibnamefont
  {Bechler}}\ and\ \bibinfo {author} {\bibfnamefont {J.}~\bibnamefont
  {Masell}},\ }\href {\doibase 10.1088/2634-4386/ace549} {\bibfield  {journal}
  {\bibinfo  {journal} {Neuromorphic Computing and Engineering}\ }\textbf
  {\bibinfo {volume} {3}},\ \bibinfo {pages} {034003}}\BibitemShut {NoStop}%
\bibitem [{\citenamefont {Filippou}\ \emph {et~al.}(2018)\citenamefont
  {Filippou}, \citenamefont {Jeong}, \citenamefont {Ferrante}, \citenamefont
  {Yang}, \citenamefont {Topuria}, \citenamefont {Samant},\ and\ \citenamefont
  {Parkin}}]{parkin_chiralDW_2018}%
  \BibitemOpen
  \bibfield  {author} {\bibinfo {author} {\bibfnamefont {P.~C.}\ \bibnamefont
  {Filippou}}, \bibinfo {author} {\bibfnamefont {J.}~\bibnamefont {Jeong}},
  \bibinfo {author} {\bibfnamefont {Y.}~\bibnamefont {Ferrante}}, \bibinfo
  {author} {\bibfnamefont {S.-H.}\ \bibnamefont {Yang}}, \bibinfo {author}
  {\bibfnamefont {T.}~\bibnamefont {Topuria}}, \bibinfo {author} {\bibfnamefont
  {M.~G.}\ \bibnamefont {Samant}}, \ and\ \bibinfo {author} {\bibfnamefont
  {S.~S.~P.}\ \bibnamefont {Parkin}},\ }\href {\doibase
  10.1038/s41467-018-07091-3} {\bibfield  {journal} {\bibinfo  {journal}
  {Nature Communications}\ }\textbf {\bibinfo {volume} {9}},\ \bibinfo {pages}
  {4653} (\bibinfo {year} {2018})}\BibitemShut {NoStop}%
\bibitem [{Note1()}]{Note1}%
  \BibitemOpen
  \bibinfo {note} {See Supplemental Material at [URL will be inserted by
  publisher]}\BibitemShut {NoStop}%
\bibitem [{Note2()}]{Note2}%
  \BibitemOpen
  \bibinfo {note} {$v_f = 1.5 \times 10^6 m/s= \protect \frac {a_0 t}{\hbar } =
  \protect \frac {1 nm \times 1 eV}{\hbar }$. This is used for comparison with
  the tight-binding model. In real MWSMs, this is roughly an order of magnitude
  lower \cite {yamanouchi_srruo3_THT_2022}, but this does not change the
  underlying physics.}\BibitemShut {Stop}%
\bibitem [{Note3()}]{Note3}%
  \BibitemOpen
  \bibinfo {note} {See Supplemental Material at [URL will be inserted by
  publisher] for full derivation.}\BibitemShut {Stop}%
\bibitem [{Note4()}]{Note4}%
  \BibitemOpen
  \bibinfo {note} {See Supplemental Material at [URL will be inserted by
  publisher] for full derivation}\BibitemShut {NoStop}%
\bibitem [{\citenamefont {Araki}\ and\ \citenamefont
  {Nomura}(2018)}]{araki2018}%
  \BibitemOpen
  \bibfield  {author} {\bibinfo {author} {\bibfnamefont {Y.}~\bibnamefont
  {Araki}}\ and\ \bibinfo {author} {\bibfnamefont {K.}~\bibnamefont {Nomura}},\
  }\href {\doibase 10.1103/PhysRevApplied.10.014007} {\bibfield  {journal}
  {\bibinfo  {journal} {Phys. Rev. Appl.}\ }\textbf {\bibinfo {volume} {10}},\
  \bibinfo {pages} {014007} (\bibinfo {year} {2018})}\BibitemShut {NoStop}%
\bibitem [{Note5()}]{Note5}%
  \BibitemOpen
  \bibinfo {note} {It is necessary to include a k-dependent mass term in order
  to create a single pair of Weyl cones in the tight-binding model, though
  $m(k) \propto (1-cos(ka))$ is near-zero in the region of the Weyl
  points}\BibitemShut {NoStop}%
\bibitem [{\citenamefont {O'Handley}(1999)}]{ohandley_modern_1999}%
  \BibitemOpen
  \bibfield  {author} {\bibinfo {author} {\bibfnamefont {R.}~\bibnamefont
  {O'Handley}},\ }\href@noop {} {\emph {\bibinfo {title} {Modern {Magnetic}
  {Materials}: {Principles} and {Applications}}}}\ (\bibinfo  {publisher}
  {Wiley},\ \bibinfo {year} {1999})\BibitemShut {NoStop}%
\bibitem [{Note6()}]{Note6}%
  \BibitemOpen
  \bibinfo {note} {We take a simple, spin and orbital-degenerate toy metal
  Hamiltonian $\protect \hat {H}_{\protect \mathrm {site}\protect \tmspace
  +\thickmuskip {.2777em}i} = \mathop {\DOTSB \sum@ \slimits@ _{d\in
  \{+,-\},j\in \{x,y,z\}}}{}\ [t c_{i+d\protect \hat {j}}^+c_i \tau _0\otimes
  \sigma _0] + (3 t - \varepsilon _0) [c_i^+c_i\tau _0\otimes \sigma _0]$ with
  an arbitrary $t = 1$ eV and $\varepsilon _0 = 1$ eV}\BibitemShut {NoStop}%
\bibitem [{Note7()}]{Note7}%
  \BibitemOpen
  \bibinfo {note} {The semi-infinite MWSM electrodes are useful in the large
  scattering limit, where an electron will locally see an infinite lattice in
  all directions. They are also most useful for resolving the Fermi-surfaces,
  where surface states will confound the fermi surface twisting}\BibitemShut
  {NoStop}%
\bibitem [{\citenamefont {Sancho}\ \emph {et~al.}(1985)\citenamefont {Sancho},
  \citenamefont {Sancho}, \citenamefont {Sancho},\ and\ \citenamefont
  {Rubio}}]{sancho_highly_1985}%
  \BibitemOpen
  \bibfield  {author} {\bibinfo {author} {\bibfnamefont {M.~P.~L.}\
  \bibnamefont {Sancho}}, \bibinfo {author} {\bibfnamefont {J.~M.~L.}\
  \bibnamefont {Sancho}}, \bibinfo {author} {\bibfnamefont {J.~M.~L.}\
  \bibnamefont {Sancho}}, \ and\ \bibinfo {author} {\bibfnamefont
  {J.}~\bibnamefont {Rubio}},\ }\href {\doibase 10.1088/0305-4608/15/4/009}
  {\bibfield  {journal} {\bibinfo  {journal} {Journal of Physics F: Metal
  Physics}\ }\textbf {\bibinfo {volume} {15}},\ \bibinfo {pages} {851}
  (\bibinfo {year} {1985})}\BibitemShut {NoStop}%
\bibitem [{\citenamefont {Breitkreiz}\ and\ \citenamefont
  {Brouwer}()}]{breitkreiz_fermi_arc_2023}%
  \BibitemOpen
  \bibfield  {author} {\bibinfo {author} {\bibfnamefont {M.}~\bibnamefont
  {Breitkreiz}}\ and\ \bibinfo {author} {\bibfnamefont {P.~W.}\ \bibnamefont
  {Brouwer}},\ }\href {\doibase 10.1103/PhysRevLett.130.196602} {\bibfield
  {journal} {\bibinfo  {journal} {Physical Review Letters}\ }\textbf {\bibinfo
  {volume} {130}},\ \bibinfo {pages} {196602}}\BibitemShut {NoStop}%
\bibitem [{Note8()}]{Note8}%
  \BibitemOpen
  \bibinfo {note} {See \cite {camsari_nonequilibrium_2023}; $G^R_{k_{||}} =
  (E+i\eta - H_{k_{||}} - \Sigma _{R,k_{||}} - \Sigma _{L,k_{||}})^{-1}; \Gamma
  _i = i(\Sigma _i - \Sigma _i^\dagger ); T = tr(\Gamma _L G^R \Gamma _R G^A);
  G^A = (G^R)^\dagger $}\BibitemShut {NoStop}%
\bibitem [{\citenamefont {Philip}\ \emph {et~al.}(2017)\citenamefont {Philip},
  \citenamefont {Hirsbrunner}, \citenamefont {Park},\ and\ \citenamefont
  {Gilbert}}]{gilbert_TI_interconnects_2017}%
  \BibitemOpen
  \bibfield  {author} {\bibinfo {author} {\bibfnamefont {T.~M.}\ \bibnamefont
  {Philip}}, \bibinfo {author} {\bibfnamefont {M.~R.}\ \bibnamefont
  {Hirsbrunner}}, \bibinfo {author} {\bibfnamefont {M.~J.}\ \bibnamefont
  {Park}}, \ and\ \bibinfo {author} {\bibfnamefont {M.~J.}\ \bibnamefont
  {Gilbert}},\ }\href {\doibase 10.1109/LED.2016.2629760} {\bibfield  {journal}
  {\bibinfo  {journal} {IEEE Electron Device Letters}\ }\textbf {\bibinfo
  {volume} {38}},\ \bibinfo {pages} {138} (\bibinfo {year} {2017})}\BibitemShut
  {NoStop}%
\bibitem [{\citenamefont {Kharzeev}\ and\ \citenamefont
  {Yee}(2013)}]{kharzeev_chiralanomaly_2013}%
  \BibitemOpen
  \bibfield  {author} {\bibinfo {author} {\bibfnamefont {D.~E.}\ \bibnamefont
  {Kharzeev}}\ and\ \bibinfo {author} {\bibfnamefont {H.-U.}\ \bibnamefont
  {Yee}},\ }\href {\doibase 10.1103/PhysRevB.88.115119} {\bibfield  {journal}
  {\bibinfo  {journal} {Phys. Rev. B}\ }\textbf {\bibinfo {volume} {88}},\
  \bibinfo {pages} {115119} (\bibinfo {year} {2013})}\BibitemShut {NoStop}%
\bibitem [{\citenamefont {Kammhuber}\ \emph {et~al.}(2017)\citenamefont
  {Kammhuber}, \citenamefont {Cassidy}, \citenamefont {Pei}, \citenamefont
  {Nowak}, \citenamefont {Vuik}, \citenamefont {Guel}, \citenamefont {Car},
  \citenamefont {Plissard}, \citenamefont {Bakkers}, \citenamefont {Wimmer},\
  and\ \citenamefont {Kouwenhoven}}]{kammhuber_helical_2017}%
  \BibitemOpen
  \bibfield  {author} {\bibinfo {author} {\bibfnamefont {J.}~\bibnamefont
  {Kammhuber}}, \bibinfo {author} {\bibfnamefont {M.~C.}\ \bibnamefont
  {Cassidy}}, \bibinfo {author} {\bibfnamefont {F.}~\bibnamefont {Pei}},
  \bibinfo {author} {\bibfnamefont {M.~P.}\ \bibnamefont {Nowak}}, \bibinfo
  {author} {\bibfnamefont {A.}~\bibnamefont {Vuik}}, \bibinfo {author}
  {\bibfnamefont {O.}~\bibnamefont {Guel}}, \bibinfo {author} {\bibfnamefont
  {D.}~\bibnamefont {Car}}, \bibinfo {author} {\bibfnamefont {S.~R.}\
  \bibnamefont {Plissard}}, \bibinfo {author} {\bibfnamefont {E.~P. a.~M.}\
  \bibnamefont {Bakkers}}, \bibinfo {author} {\bibfnamefont {M.}~\bibnamefont
  {Wimmer}}, \ and\ \bibinfo {author} {\bibfnamefont {L.~P.}\ \bibnamefont
  {Kouwenhoven}},\ }\href {\doibase 10.1038/s41467-017-00315-y} {\bibfield
  {journal} {\bibinfo  {journal} {Nature Communications}\ }\textbf {\bibinfo
  {volume} {8}},\ \bibinfo {pages} {478} (\bibinfo {year} {2017})}\BibitemShut
  {NoStop}%
\bibitem [{\citenamefont {Wolf}\ \emph {et~al.}(2021)\citenamefont {Wolf},
  \citenamefont {Zilberberg}, \citenamefont {Blatter},\ and\ \citenamefont
  {Lado}}]{wolf_periodicgraphene_2021}%
  \BibitemOpen
  \bibfield  {author} {\bibinfo {author} {\bibfnamefont {T.~M.}\ \bibnamefont
  {Wolf}}, \bibinfo {author} {\bibfnamefont {O.}~\bibnamefont {Zilberberg}},
  \bibinfo {author} {\bibfnamefont {G.}~\bibnamefont {Blatter}}, \ and\
  \bibinfo {author} {\bibfnamefont {J.~L.}\ \bibnamefont {Lado}},\ }\href
  {\doibase 10.1103/PhysRevLett.126.056803} {\bibfield  {journal} {\bibinfo
  {journal} {Physical Review Letters}\ }\textbf {\bibinfo {volume} {126}},\
  \bibinfo {pages} {056803} (\bibinfo {year} {2021})}\BibitemShut {NoStop}%
\bibitem [{\citenamefont {Paul}\ \emph {et~al.}(2023)\citenamefont {Paul},
  \citenamefont {Zhang},\ and\ \citenamefont {Fu}}]{paul_skyrmionlattice_2023}%
  \BibitemOpen
  \bibfield  {author} {\bibinfo {author} {\bibfnamefont {N.}~\bibnamefont
  {Paul}}, \bibinfo {author} {\bibfnamefont {Y.}~\bibnamefont {Zhang}}, \ and\
  \bibinfo {author} {\bibfnamefont {L.}~\bibnamefont {Fu}},\ }\href {\doibase
  10.1126/sciadv.abn1401} {\bibfield  {journal} {\bibinfo  {journal} {Science
  Advances}\ }\textbf {\bibinfo {volume} {9}},\ \bibinfo {pages} {eabn1401}
  (\bibinfo {year} {2023})}\BibitemShut {NoStop}%
\bibitem [{\citenamefont {Muduli}\ \emph {et~al.}(2019)\citenamefont {Muduli},
  \citenamefont {Higo}, \citenamefont {Nishikawa}, \citenamefont {Qu},
  \citenamefont {Isshiki}, \citenamefont {Kondou}, \citenamefont
  {Nishio-Hamane}, \citenamefont {Nakatsuji},\ and\ \citenamefont
  {Otani}}]{muduli_mn3sn_2019}%
  \BibitemOpen
  \bibfield  {author} {\bibinfo {author} {\bibfnamefont {P.~K.}\ \bibnamefont
  {Muduli}}, \bibinfo {author} {\bibfnamefont {T.}~\bibnamefont {Higo}},
  \bibinfo {author} {\bibfnamefont {T.}~\bibnamefont {Nishikawa}}, \bibinfo
  {author} {\bibfnamefont {D.}~\bibnamefont {Qu}}, \bibinfo {author}
  {\bibfnamefont {H.}~\bibnamefont {Isshiki}}, \bibinfo {author} {\bibfnamefont
  {K.}~\bibnamefont {Kondou}}, \bibinfo {author} {\bibfnamefont
  {D.}~\bibnamefont {Nishio-Hamane}}, \bibinfo {author} {\bibfnamefont
  {S.}~\bibnamefont {Nakatsuji}}, \ and\ \bibinfo {author} {\bibfnamefont
  {Y.}~\bibnamefont {Otani}},\ }\href {\doibase 10.1103/PhysRevB.99.184425}
  {\bibfield  {journal} {\bibinfo  {journal} {Physical Review B}\ }\textbf
  {\bibinfo {volume} {99}},\ \bibinfo {pages} {184425} (\bibinfo {year}
  {2019})}\BibitemShut {NoStop}%
\bibitem [{\citenamefont {Saito}\ and\ \citenamefont
  {Nishio-Hamane}(2021)}]{saito_co2mnga_2021}%
  \BibitemOpen
  \bibfield  {author} {\bibinfo {author} {\bibfnamefont {T.}~\bibnamefont
  {Saito}}\ and\ \bibinfo {author} {\bibfnamefont {D.}~\bibnamefont
  {Nishio-Hamane}},\ }\href {\doibase 10.1016/j.physb.2020.412761} {\bibfield
  {journal} {\bibinfo  {journal} {Physica B: Condensed Matter}\ }\textbf
  {\bibinfo {volume} {603}},\ \bibinfo {pages} {412761} (\bibinfo {year}
  {2021})}\BibitemShut {NoStop}%
\bibitem [{\citenamefont {Yao}\ \emph {et~al.}(2018)\citenamefont {Yao},
  \citenamefont {Lee}, \citenamefont {Xu}, \citenamefont {Wang}, \citenamefont
  {Ma}, \citenamefont {Yazyev}, \citenamefont {Xiong}, \citenamefont {Shi},
  \citenamefont {Aeppli},\ and\ \citenamefont {Soh}}]{yao_fe3sn2_2018}%
  \BibitemOpen
  \bibfield  {author} {\bibinfo {author} {\bibfnamefont {M.}~\bibnamefont
  {Yao}}, \bibinfo {author} {\bibfnamefont {H.}~\bibnamefont {Lee}}, \bibinfo
  {author} {\bibfnamefont {N.}~\bibnamefont {Xu}}, \bibinfo {author}
  {\bibfnamefont {Y.}~\bibnamefont {Wang}}, \bibinfo {author} {\bibfnamefont
  {J.}~\bibnamefont {Ma}}, \bibinfo {author} {\bibfnamefont {O.~V.}\
  \bibnamefont {Yazyev}}, \bibinfo {author} {\bibfnamefont {Y.}~\bibnamefont
  {Xiong}}, \bibinfo {author} {\bibfnamefont {M.}~\bibnamefont {Shi}}, \bibinfo
  {author} {\bibfnamefont {G.}~\bibnamefont {Aeppli}}, \ and\ \bibinfo {author}
  {\bibfnamefont {Y.}~\bibnamefont {Soh}},\ }\href {\doibase
  10.48550/arXiv.1810.01514} {} (\bibinfo {year} {2018}),\ \Eprint
  {http://arxiv.org/abs/1810.01514 [cond-mat]} {1810.01514 [cond-mat]}
  \BibitemShut {NoStop}%
\bibitem [{\citenamefont {Sorn}\ and\ \citenamefont
  {Paramekanti}(2021)}]{sorn_skewscattering_2021}%
  \BibitemOpen
  \bibfield  {author} {\bibinfo {author} {\bibfnamefont {S.}~\bibnamefont
  {Sorn}}\ and\ \bibinfo {author} {\bibfnamefont {A.}~\bibnamefont
  {Paramekanti}},\ }\href {\doibase 10.1103/PhysRevB.103.104413} {\bibfield
  {journal} {\bibinfo  {journal} {Physical Review B}\ }\textbf {\bibinfo
  {volume} {103}},\ \bibinfo {pages} {104413} (\bibinfo {year}
  {2021})}\BibitemShut {NoStop}%
\bibitem [{\citenamefont {Xuan~Mei}\ \emph {et~al.}(2021)\citenamefont
  {Xuan~Mei}, \citenamefont {Chen},\ and\ \citenamefont
  {Li}}]{xuan_tiltedMR_2021}%
  \BibitemOpen
  \bibfield  {author} {\bibinfo {author} {\bibfnamefont {X.}~\bibnamefont
  {Xuan~Mei}}, \bibinfo {author} {\bibfnamefont {M.}~\bibnamefont {Chen}}, \
  and\ \bibinfo {author} {\bibfnamefont {H.}~\bibnamefont {Li}},\ }\href
  {https://aip.scitation.org/doi/abs/10.1063/5.0073023} {\bibfield  {journal}
  {\bibinfo  {journal} {Journal of Applied Physics}\ }\textbf {\bibinfo
  {volume} {130}},\ \bibinfo {pages} {203901} (\bibinfo {year}
  {2021})}\BibitemShut {NoStop}%
\bibitem [{\citenamefont {Leonard}\ \emph {et~al.}(2022)\citenamefont
  {Leonard}, \citenamefont {Liu}, \citenamefont {Alamdar}, \citenamefont {Jin},
  \citenamefont {Cui}, \citenamefont {Akinola}, \citenamefont {Xue},
  \citenamefont {Xiao}, \citenamefont {Friedman}, \citenamefont {Marinella},
  \citenamefont {Bennett},\ and\ \citenamefont
  {Incorvia}}]{leonard_synapses_2022}%
  \BibitemOpen
  \bibfield  {author} {\bibinfo {author} {\bibfnamefont {T.}~\bibnamefont
  {Leonard}}, \bibinfo {author} {\bibfnamefont {S.}~\bibnamefont {Liu}},
  \bibinfo {author} {\bibfnamefont {M.}~\bibnamefont {Alamdar}}, \bibinfo
  {author} {\bibfnamefont {H.}~\bibnamefont {Jin}}, \bibinfo {author}
  {\bibfnamefont {C.}~\bibnamefont {Cui}}, \bibinfo {author} {\bibfnamefont
  {O.~G.}\ \bibnamefont {Akinola}}, \bibinfo {author} {\bibfnamefont
  {L.}~\bibnamefont {Xue}}, \bibinfo {author} {\bibfnamefont {T.~P.}\
  \bibnamefont {Xiao}}, \bibinfo {author} {\bibfnamefont {J.~S.}\ \bibnamefont
  {Friedman}}, \bibinfo {author} {\bibfnamefont {M.~J.}\ \bibnamefont
  {Marinella}}, \bibinfo {author} {\bibfnamefont {C.~H.}\ \bibnamefont
  {Bennett}}, \ and\ \bibinfo {author} {\bibfnamefont {J.~A.~C.}\ \bibnamefont
  {Incorvia}},\ }\href {\doibase 10.1002/aelm.202200563} {\bibfield  {journal}
  {\bibinfo  {journal} {Advanced Electronic Materials}\ ,\ \bibinfo {pages}
  {2200563}} (\bibinfo {year} {2022})}\BibitemShut {NoStop}%
\bibitem [{\citenamefont {Camsari}\ \emph {et~al.}(2023)\citenamefont
  {Camsari}, \citenamefont {Chowdhury},\ and\ \citenamefont
  {Datta}}]{camsari_nonequilibrium_2023}%
  \BibitemOpen
  \bibfield  {author} {\bibinfo {author} {\bibfnamefont {K.~Y.}\ \bibnamefont
  {Camsari}}, \bibinfo {author} {\bibfnamefont {S.}~\bibnamefont {Chowdhury}},
  \ and\ \bibinfo {author} {\bibfnamefont {S.}~\bibnamefont {Datta}},\ }in\
  \href {\doibase 10.1007/978-3-030-79827-7_44} {\emph {\bibinfo {booktitle}
  {Springer {Handbook} of {Semiconductor} {Devices}}}},\ \bibinfo {series and
  number} {Springer {Handbooks}},\ \bibinfo {editor} {edited by\ \bibinfo
  {editor} {\bibfnamefont {M.}~\bibnamefont {Rudan}}, \bibinfo {editor}
  {\bibfnamefont {R.}~\bibnamefont {Brunetti}}, \ and\ \bibinfo {editor}
  {\bibfnamefont {S.}~\bibnamefont {Reggiani}}}\ (\bibinfo  {publisher}
  {Springer International Publishing},\ \bibinfo {address} {Cham},\ \bibinfo
  {year} {2023})\ pp.\ \bibinfo {pages} {1583--1599}\BibitemShut {NoStop}%
\end{thebibliography}%
